\documentclass[11pt]{article}

\usepackage{amscd }
\usepackage{amsfonts}
\usepackage{graphicx}
\usepackage{fullpage}
\usepackage{wrapfig,lipsum} 
\usepackage{epsfig,psfrag,amsmath,amssymb,latexsym}
\usepackage{amscd}
\usepackage{xcolor}
\usepackage{float}
\usepackage{algorithmicx, algpseudocode}
\usepackage{algorithm}
\usepackage{amsthm, cases, hyperref, mathtools, url}
\usepackage{natbib}
\usepackage{subcaption}
\usepackage{authblk}
\usepackage{hyperref}

\newcommand{\hide}[1]{}
\newcommand*{\Reals}{\mathbb{R}} 
\newcommand{\set}[1]{\left\{ \, #1 \, \right\} }
\newcommand*{\bm}[1]{{\bf #1}}

\newcommand*{\G}{\bm{G}}

\newcommand{\an}[1]{\begin{align}#1\end{align}}	
\newcommand{\ab}[1]{\begin{align*}#1\end{align*}}	

\newcommand{\dsty}{\displaystyle}
\newcommand*{\eqdef}{\, {:=}\, }

\newcommand*{\ov}{\overline}
\newcommand*{\xb}{\ov{x}}

\newcommand*{\phih}{\hat{\phi}}

\newcommand*{\pf}[2]{\displaystyle \frac{\partial #1}{\partial #2}}
\newcommand{\Cases}[1]{\left\{ \begin{array}{ll} \displaystyle #1 \end{array}\right.}   
 
\newcommand{\alphah}{\hat{\alpha}}
\newcommand{\itemref}[1]{\ref{#1}.\!\!}
\newcommand{\Quote}[1]{\begin{quote}#1\end{quote}}

\newcommand*{\tr}{^{\!\top}}	
\newcommand*{\xh}{{\hat{x}}}
\newcommand*{\fracinline}[2]{{#1 / #2}}
\newcommand*{\x}{\bm{x}}

\newcommand*{\y}{\bm{y}}
\newcommand*{\z}{\bm{z}}

\newcommand*{\Prob}{{\sf Prob}}

\newcommand*{\moddd}{\ {\sf mod}\,}
\newcommand*{\Normal}{{\cal N}}
\newcommand*{\wb}{\overline{w}}
\newcommand*{\Var}{{\sf Var}} 
\newcommand*{\0}{{\bm 0}}
\newcommand*{\I}{\bm{I}}

\begin{document}

\title{The Evolution of Imitation \\Without Cultural Transmission}

\author[1]{Lee Altenberg\footnote{\href{mailto:altenber@hawaii.edu?subject=Evolution_of_Imitation}{altenber@hawaii.edu}}}
\author[1]{ Susanne Still}
\author[2]{Christopher J. Watkins}
\affil[1]{University of Hawai`i at M\=anoa}
\affil[2]{Royal Holloway University of London}
\date{\today}
\maketitle

\begin{abstract}
The evolution and function of imitation have always been placed within the confines of animal learning and associated with its crucial role in cultural transmission and cultural evolution.  Can imitation evolve as a form of phenotypic plasticity in the absence of cultural transmission, in phenotypes beyond behavior?  

We investigate a model in which imitation is unbundled from cultural transmission:  an organism's adult phenotype is plastically altered by its experiences as a juvenile of other juveniles' genetically determined traits.  The only information transmitted between generations is genetic. 

We find that during a period of directional selection towards a phenotypic optimum, natural selection favors modifiers which cause an organism to bias its plastic phenotype in the direction \emph{opposite} to the mean phenotype of the population---\emph{anti-imitation}.  As the population approaches the phenotypic optimum and shifts into stabilizing selection, selection on the modifier reverses and favors strong imitation of the population mean.  Imitation can evolve to overshoot the target and produce an evolutionary pathology where mean fitness decreases.  When purifying selection for an extreme phenotype is modeled, only selection for anti-imitation occurs, even at a mutation-selection balance.  

Imitation and anti-imitation emerge from these models in the absence of any goal representation,  cognitive understanding of its purpose, or discernment of any kind.  These theoretical outcomes are all novel evolutionary and biological phenomena, and we discuss their implications. 
\end{abstract}

\section{Introduction}
Imitation is regarded as the core capability that enables cultural transmission and evolution, and conversely, cultural transmission is regarded as the principal function of imitation.  Imitation can be considered as a form of phenotypic plasticity in which the development of an organism's phenotype is sensitive to the phenotypes of other organisms.  

Imitation, in order to evolve as an adaptation, needs to confer a selective advantage upon the imitator.  The question arises as to why imitating the states of other organism would be adaptive.  The classical proposed mechanisms are (1) the imitator can discern which individuals or behaviors to imitate so as to choose the more adaptive among them (\emph{selective transmission}), or (2) the conspecifics who are imitated have undergone individual learning in their development, which makes their behaviors more adaptive than the imitator's default state (\emph{adaptive learning}), and thus imitation can avoid costs of trial-and-error learning \citep{Cavalli-Sforza:and:Feldman:1973:Cultural,Richerson:and:Boyd:1978:Dual,Cavalli-Sforza:and:Feldman:1983:Cultural}. 

This nearly universal bundling of imitation with the other processes of selective transmission and individual learning puts it in a box which limits full exploration of exactly what imitation is capable of doing in evolutionary dynamics.  A way to develop a more full understanding of imitation is to go outside the box and break it out of this bundle.  \citet{Gonzalez:Watson:and:Bullock:2017:Minimally} have initiated this unbundling by asking whether selective transmission and  adaptive learning are necessary for the evolution of imitation.  They investigate whether imitation could evolve in the absence of selective transmission and adaptive learning through a model where (1) imitators choose without bias the model individual to imitate, and (2) their imitated phenotype remains unimproved through any kind of learning during their lifetime.  Under specific population dynamic conditions (a Moran model with cultural transmission and strong viability selection) imitation evolves with high likelihood in the absence of selective transmission or adaptive learning.  Additionally, they find that when  cultural transmission evolves, it irreversibly displaces genetic information as the dominant mode of inheritance. 

In the current work, we assume that there is neither selective transmission nor adaptive individual learning, as \citet{Gonzalez:Watson:and:Bullock:2017:Minimally} have done.  We further unbundle imitation by removing cultural transmission from the dynamics.  Cultural transmission occurs in the \citet{Gonzalez:Watson:and:Bullock:2017:Minimally} model through \emph{oblique transmission} \citep{Cavalli-Sforza:and:Feldman:1981} in which juveniles imitate the phenotypes of adults and then exhibit phenotypes which can \emph{themselves be imitated} by later generations.  This makes the adult phenotype a channel of transmission separate from the genotype.  Here, we strip away cultural transmission by preventing any imitation of phenotypes that were themselves imitations.  Imitation is restricted to one-shot sampling of genetically determined phenotypes of juveniles by juveniles.  The adult phenotypes that result from imitation cannot themselves be further imitated; it is thus ``acquired variation'' \citep{Boyd:and:Richerson:1983:Cultural}, but not culturally transmitted.  This eliminates the phenotype as a pool of transmitted information separate from the genotype.  The only communication between generations is genetic transmission.   

We suppose that organisms have a quantitative genetic trait which is expressed early in development.  Juvenile organisms gather sensory information about the quantitative trait as expressed by the juveniles in their cohort.  A \emph{phenotypic plasticity function} then combines the organism's own juvenile quantitative trait with this population information to determine the organism's adult phenotype.  Natural selection acts only on this adult phenotype after all plastic development has happened.  To prevent any cultural transmission between generations, we model a semelparous species with discrete, non-overlapping generations, where the parental generation is gone and any information about their plastic phenotypes erased by the time offspring develop.    

Our focal question is the evolution of this phenotypic plasticity function.  We model a modifier gene that creates genetic variation over a space of plasticity functions.  For a particular plasticity function to be adaptive, it must somehow take the information it has available --- solely the distribution of genetically determined juvenile phenotypes in the population --- and convert this information into modification of an organism's development so that its fitness is increased, thereby increasing the frequency of the modifier gene variant.

\subsection{Additional Clarifications}
A common term used almost synonymously with imitation is \emph{social learning}, where the focal phenotype is behavior.  Because there is nothing in our treatment that restricts the phenotype to behavior, we will not speak of social learning specifically here, but it should be understood that it is included within the more general category of phenotypic plasticity.  It should also be understood that under the umbrella of ``imitation'' we include \emph{copying} and \emph{emulation}   \citep{Charbonneau:and:Bourrat:2021:Fidelity}.  They point out that these distinctions depend on the level of course-graining applied to behavior.  Our simple model here considers the phenotype at the most coarse-grained level where no inner structure is distinguished.  We make no assumptions at all about the perceptual or cognitive level at which imitation occurs.

Imitation in our model is a form of \emph{horizontal transmission} \citep{Cavalli-Sforza:and:Feldman:1981} (Matthew Zefferman, personal communication) in that juveniles collect information on other organisms in their generational cohort, and this information affects their phenotype.  We preclude this phenotypic information from being transmitted to future generations in order that it not produce cultural evolution.  This separation of horizontal transmission from its usual context of cultural transmission is therefore another unbundling to be found in our model.

Lastly, we do not provide imitation as a ready-made form of genetic variation.  To do so would be to create a form of mutational bias \citep{Yampolsky:and:Stoltzfus:2001} that could artifactually boost the evolvability of the trait.  Imitation in nature involves multiple chains of information processing from sensory systems to effector systems (with the possible exception of prions, in which folding could be considered to be direct imitation).  For a single mutation to produce an offspring with fully formed imitation from a parent with no imitation seems biologically unlikely.  Instead, we here assume that:
\begin{enumerate}
\item  an organism receives a perceptual signal about the phenotypes of its cohort;
\item this perceived signal is used by the organism's developmental system to plastically shape its adult phenotype,
\item this plastic shaping is drawn from a set of genetically determined {phenotypic plasticity functions}; and
\item mutation in the space of these possible functions is incremental and not biased toward imitation.
\end{enumerate}
The phenotypic plasticity function is modeled as a quantitative trait parameter that may range over a continuum going from no effect, to imitation, to what we call ``anti-imitation'', in which the perceptual signal received by the organism causes its phenotype to develop in the \emph{opposite} direction the phenotype that produced the signal.

It should be clarified here that anti-imitation is not analogous to \emph{anti-conformism} \citep{Boyd:and:Richerson:1985,Liberman:Ram:Altenberg:and:Feldman:2020,Denton:etal:2020:Cultural}.  Conformism and anti-conformism are defined as forms of frequency-dependent transmission bias, but what is transmitted is always an imitation of the model.  Anti-imitation is orthogonal to transmission bias, in that it is the phenotypic \emph{response} to a model, not the \emph{choice} of the model.   Anti-conformism is a form of imitation where there is transmission bias toward imitating less-frequent phenotypes.  Anti-imitation, in contrast, is a response of the organism to a model wherein it plastically alters its metric phenotype in a direction \emph{opposite} to that of the model.

It should be noted that other concepts referred to as ``anti-imitation'' have appeared in several lines of research on sequential multi-player games of partial information \citep{Eyster:and:Rabin:2010:Naive,Dasaratha:Golub:and:Hak:2020:Learning,Kendall:2021:Herding,Zhang:2021:Non-monotone}.  Their usage is related to dynamics such as herding and contrarianism and the problem of redundant information.  Again these are distinct from our usage.

In a sense, the task of the plasticity function is to act like a population geneticist --- to take a snapshot of the genetic variation in the population and make inferences from this information about the shape of the adaptive landscape, specifically, the direction of a phenotypic optimum.  The idea that a snapshot of the genetic variation in the population can give an organism information which it can use adaptively in its development is an instance of the idea of ``genes-as-cues'' \citep{Leimar:2005:Evolution,Leimar:Hammerstein:and:VanDooren:2006}. From \citet{Dall:McNamara:and:Leimar:2015:Genes}:
\Quote{
Adaptive cue-integration systems of the sort outlined
above will take any available sources of information (epigenetic
and genetic) into account in determining phenotypes.
}

\subsection{Our Contributions}
Our chief findings are the following: 
\begin{enumerate}
\item During directional selection toward a phenotypic optimum, \emph{anti-imitation} is strongly selected for, and it amplifies adult phenotypic and fitness variance.
\item As directional selection shifts into stabilizing selection at a mutation-selection balance around a phenotypic optimum, strong selection for imitation takes over, and imitation suppresses adult phenotypic and fitness variance.
\item Selection for imitation during stabilizing selection can reach such an extreme that \emph{genotype-phenotype disengagement} occurs \citep{Gonzalez:Watson:and:Bullock:2017:Minimally}, in which the  the organism's own genotype no longer influences its adult phenotype. \label{item:GPD} 
\item Stabilizing selection can further push the population into \emph{hyper-imitation}.  This drives the phenotype away from the optimum to such a point that the population is again under directional selection, and imitation levels collapse.  This collapse phenomenon becomes very sharp in models of multivariate phenotypes.  Evolution goes through repeated cycles of hyper-imitation and imitation collapse. The precise timing of these collapses of imitation involves a great deal of randomness.
\item If genetic constraints prevent the production of hyper-imitation variants, populations under stabilizing selection reach stable high values of imitation.
\item In a model of purifying selection for an extreme trait, where phenotypic variation is restricted to one side of the optimum, very strong anti-imitation is selected for even at a mutation-selection balance.
\end{enumerate}
We are not aware of any of these phenomena save \itemref{item:GPD} having been documented nor hypothesized previously.  As novel phenomena, they provide new search images for their occurrence empirical studies.

\section{The Model}

We now specify the details of the model investigated here: the organism, the phenotypic plasticity function, fitnesses, and the population dynamics.

\subsection{The Organism}

Organisms in this model species contain 2 pieces of inherited information: 
\begin{enumerate}
\item major loci that specify the juvenile phenotype,
\item a modifier locus that specifies the phenotypic plasticity function.
\end{enumerate}

The genotype of the major loci determines the phenotype of the juvenile organism.  The juvenile organism observes the other juvenile phenotypes in the population, and these observations combined with its own juvenile phenotype are inputs into its phenotypic plasticity function.   Its particular plasticity function is determined by its modifier locus genotype.  The organism then develops into an adult whose phenotype is the output of its plasticity function.  The fitness of the organism is determined by its adult phenotype.

We assume the organisms are haploid and reproduce asexually.  During reproduction, both the major loci and the modifier locus are subject to mutation.  We use two models for the structure of these genotypes:  
\begin{enumerate}
\item \label{item:quant} A quantitative-genetic genotype represented by a multivariate genotypic value, $\y = (y_1, \ldots, y_D) \in \Reals^D$, where $D$ is the number of phenotype dimensions, which mutation perturbs by a normal Gaussian random variable distributed as $\Normal(0, \sigma^2)$.
\item \label{item:string} A multilocus biallelic model, $\y = (y_1, \ldots, y_L) \in \set{0,1}^L$, in which each of $L$ loci independently mutates to the alternate allele with probability $\mu$.
\end{enumerate}
The quantitative genetic representation \ref{item:quant} is appropriate for modeling stabilizing selection, while the multilocus model \ref{item:string} is appropriate for modeling purifying selection for an extreme phenotype \citep{Charlesworth:2013:Stabilizing}, which is how it is used in \citet{Gonzalez:Watson:and:Bullock:2017:Minimally}.

\begin{table}[!ht]
\begin{tabular}{llll}
\hline
$t$ && generation number \\
$N$ && population size \\
$D$ && number of phenotypic dimensions in the\\&&\qquad multivariate quantitative model \\
$L$ && number of loci for the biallelic major loci genotype \\
$\y$ &$= (y_1, \ldots, y_L) \in \set{0,1}^L$ & major loci genotype \\
$\x$ &$ = (x_1, \ldots, x_D) \in [0,1]$ or $\Reals^D$ & multivariate juvenile phenotype \\
$\xh$ && sampled juvenile phenotype value\\
$\xb$ &$=(\fracinline{1}{N}) \sum_{i=1}^N x_i$ & population mean juvenile phenotype value\\ 
$\alpha$ && phenotypic plasticity function parameter \\
$\overline{\alpha}$ &$=(\fracinline{1}{N}) \sum_{i=1}^N \alpha_i$& population mean $\alpha$ \\
$\phi$ &$=(1-\alpha) x + \alpha\, \xh$ & adult phenotype \\
$\phih$ && optimal phenotype \\
$\beta$ && selection strength parameter \\
$w(\phi)$ &$= \exp(-\beta (\phi-\phih)\tr(\phi-\phih) )$ & fitness of phenotype $\phi$ \\
$\wb$ &$ = (\fracinline{1}{N}) \sum_{i=1}^N w(\phi_i)$ & mean fitness of the population \\
\hline
\end{tabular}
\caption{Definitions of symbols used}\label{table:symbols}
\end{table}

\subsection{The Genotype-Phenotype Map}

The genotype of the major loci specifies the juvenile phenotype.  In the case of the quantitative genetic model, the juvenile phenotype $\x$ is simply the genotypic value, $\y$.  

In the multilocus case, we are interested in purifying selection on an extreme phenotype, so the juvenile phenotype is set to the fraction of $1$ alleles in the genotype:
\ab{
x(\y) &= x( (y_1, \ldots, y_L) ) = \frac{1}{L} \sum_{k=1}^L y_k \in [0,1].
}

The juvenile phenotype $x$ develops plasticly into an adult phenotype $\phi$ which determines fitness.  The plasticity function is determined by the genotypic value of the modifier locus which is described next.

\subsection{Phenotypic Plasticity Functions}
The space of possible phenotypic plasticity functions is unbounded.  In this study we restrict the space of functions to a family with a single parameter, $\alpha$, where the adult phenotype $\phi$ is chosen to fall somewhere on a line between the organism's juvenile phenotype $x$ and a \emph{sample value} $\xh$ from the juvenile phenotypes it observes in its cohort.  The parameter, $\alpha$, determines where on this line its adult phenotype falls. So
\an{\label{eq:phi}
\phi = (1-\alpha) x + \alpha \xh.
}
We describe the plasticity as \emph{imitation} when $\alpha > 0$ and \emph{anti-imitation} when $\alpha < 0$.  The case $\alpha = 0$ represents an absence of phenotypic plasticity.

We note that for $\alpha >1$, the plastic phenotype $\phi$ will overshoot the input phenotype $\xh$ and be further from $x$ that $\xh$.  We therefore refer to $\alpha > 1$ as \emph{hyper-imitation}, and its relevance will be seen in the evolutionary dynamics.

With a population of size $N$, an individual can observe in its cohort potentially $N-1$ different values of the juvenile phenotype.  In the current work, we use the population mean juvenile phenotype $\xb=(\fracinline{1}{N}) \sum_{i=1}^N x_i$ for the sample input to the plasticity function.  Other choices not explored in the current work could include where $\xh$ is the juvenile phenotype of a single randomly encountered cohort member, or a function of several encountered cohort members.

\begin{figure}[H]
\centerline{\includegraphics[width=5in]{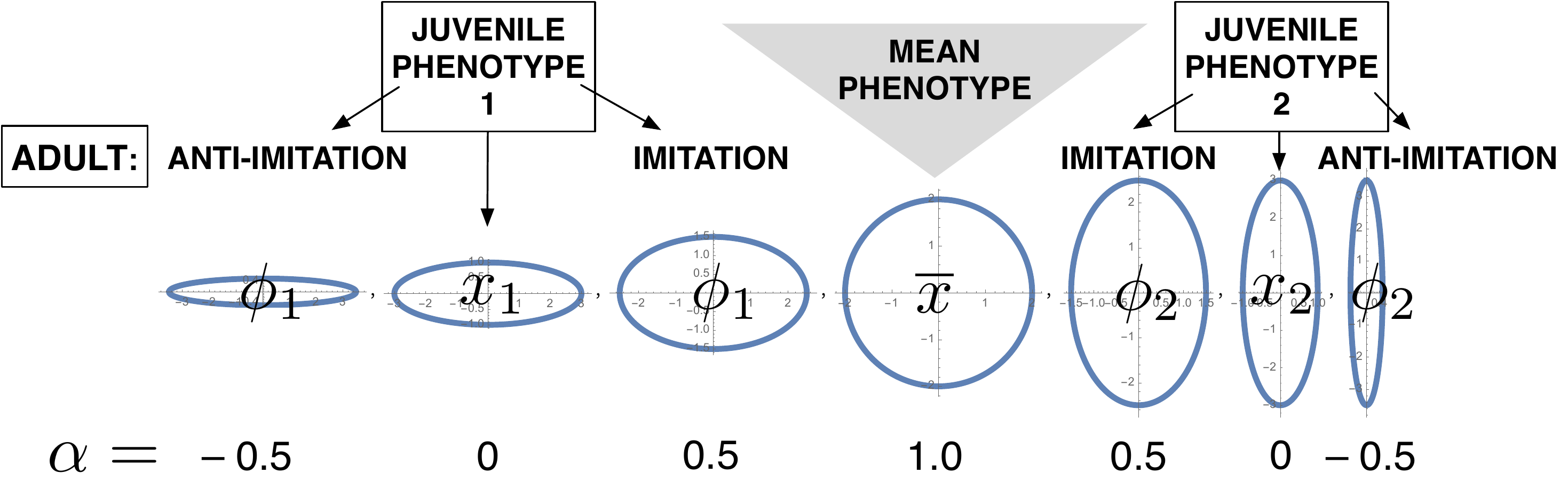}}
\caption{{\bf The phenotypic plasticity function.} Depicted for a population of two organisms.  The two juvenile phenotypes, $x_1$ and $x_2$, are shown as two ellipses with different parameters.  The population mean $\xb= (x_1+x_2)/2$ produces the circle as the mean phenotype, depicted.  Ellipses with the parameters $\phi = (1-\alpha) x + \alpha\, \xb$ are shown for $\alpha= -0.5$ (anti-imitation, pushing the adult phenotype away from the mean), $0.5$ (imitation, pushing the adult phenotype toward the mean), $0$ (no plasticity), and $1$ (all  adult phenotypes equal the mean, i.e.\ genotype-phenotype disengagement \citep{Gonzalez:Watson:and:Bullock:2017:Minimally}).  \label{fig:Example}}
\end{figure}

\begin{figure}[H]
\centerline{\includegraphics[width=5in]{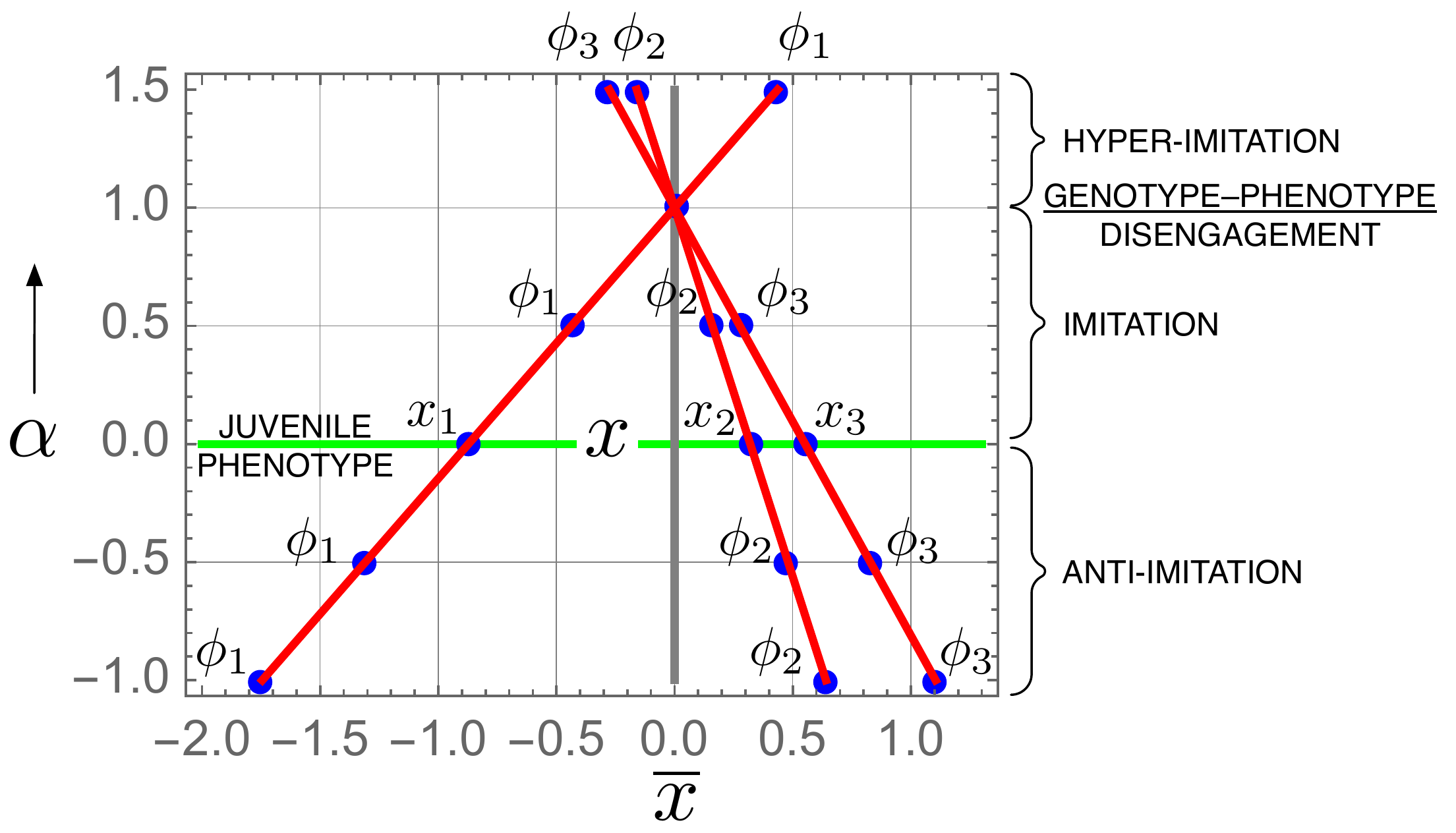}}
\caption{{\bf The effect of plasticity parameter $\alpha$ on the adult phenotype $\phi$.}  Shown are three juvenile phenotypes, $x_1, x_2, x_3$ whose values at $\alpha=0$ equal the adult phenotypes (green line).  The mean $\xb=(x_1+x_2+x_3)/3$ is set $0$.  The adult phenotypes $\phi_1, \phi_2, \phi_3$ resulting from values of $\alpha=-1, -0.5, 0.5, 1.5$ are plotted along red lines. Genotype-phenotype disengagement occurs when $\alpha=1$ at which point all adult phenotypes coincide: $\phi_1=\phi_2=\phi_3=\xb = 0$.  The ranges of $\alpha$ corresponding to anti-imitation ($\alpha<0$), imitation ($0<\alpha<1$), and hyper-imitation ($1<\alpha$) are labeled.\label{fig:AlphaGrid}}
\end{figure}

We illustrate the phenotypic plasticity function in Figures \ref{fig:Example} and \ref{fig:AlphaGrid}.  Figure \ref{fig:Example} shows a population of two organisms, with phenotypes being the shapes of ovals.  The resulting shapes for the adult phenotypes of both organisms are shown for different values of $\alpha$.  Figure  \ref{fig:AlphaGrid} shows a population of three organisms, with phenotypes depicted as the coordinates on the $X$ axis.  Again, the resulting coordinates of adult phenotypes are depicted for different values of $\alpha$ ranging from anti-imitation to genotype-phenotype disengagement to hyper-imitation.

The phenotypic plasticity parameter has a simple effect on the variance of the adult phenotype as compared to the genetically determined variance of the juvenile phenotype.  When the sample $\xh$ is set to the population mean $\xb$, the ratio of adult to juvenile phenotypic variance at any stage of evolution evaluates to:
\an{\label{eq:VarRatio}
 \frac{\Var[\phi]} {\Var[x]}
&= \frac{\Var[(1-\alpha) x + \alpha \xb]} {\Var[x]}
= \frac{(1-\alpha)^2 \, \Var[ x]} {\Var[x]}
= (1-\alpha)^2.
}
Therefore, imitation ($0 < \alpha < 1$) reduces the adult phenotypic variance, while anti-imitation ($\alpha < 0$) and large hyper-imitation ($2 < \alpha$) amplify the adult phenotypic variance.  Where there is no phenotypic plasticity ($\alpha=0$) adult phenotypes remain unchanged from juvenile phenotypes.  Genotype-phenotype disengagement ($\alpha=1$) results in there being no adult phenotypic variance.

\subsection{Fitness Functions}  
We adopt Fisher's geometric model of natural selection \citep[pp. 38--41]{Fisher:1930:Genetical}\citep{Matuszewski:Hermisson:and:Kopp:2014:Fisher}:  this assumes there is an optimal phenotype $\phih$ with maximal fitness.  The fitness is a Gaussian function of the departure of the adult phenotype $\phi$ from the optimal phenotype $\phih$:
\an{
w(\phi) &= e^{\dsty -\beta(\phi - \phih)\tr (\phi - \phih)},
}
where $\beta$ is the selection strength parameter and $\tr$ is the transpose.  Substitution of eq. \eqref {eq:phi} yields
\ab{
w(\phi(x)) &
= \exp({\dsty -\beta(\phi(x) - \phih)\tr (\phi(x) - \phih) }) \\
&= \exp({\dsty -\beta( x + \alpha(\xh - x)  - \phih)\tr ( x + \alpha(\xh - x)  - \phih) }),
}
For given values of $x$, $\xh$, and $\phi$, the optimal value $\alphah$ that maximizes fitness is obtained by differentiation:
\ab{
\pf{w(\phi )}{\alpha} &
= \pf{}{\alpha} \exp({\dsty -\beta( x + \alpha(\xh - x)  - \phih)\tr ( x + \alpha(\xh - x)  - \phih) }) \\&
= -2 \beta (\xh - x)\tr ( x + \alpha(\xh - x)  - \phih) w(\phi ).
}
hence 
\hide{
\pf{w(\phi )}{\alpha} &
= \pf{}{\alpha}  \exp({\dsty -( x + \alpha(\xh - x)  - \phih)^2 }) \\&
= -2 (\xh - x)( x + \alpha(\xh - x)  - \phih)  \exp({\dsty -( x + \alpha(\xh - x)  - \phih)^2 }),
}
$\pf{w(\phi )}{\alpha}=0$ when $x = \xh$ or when 
\an{\label{eq:alphah}
\alpha &= \hat{\alpha} \eqdef \frac{(x-\xh)\tr ( x  - \phih)}{(x-\xh)\tr (x-\xh)}.
}
\begin{figure}[H]
\centerline{\includegraphics[width=3in]{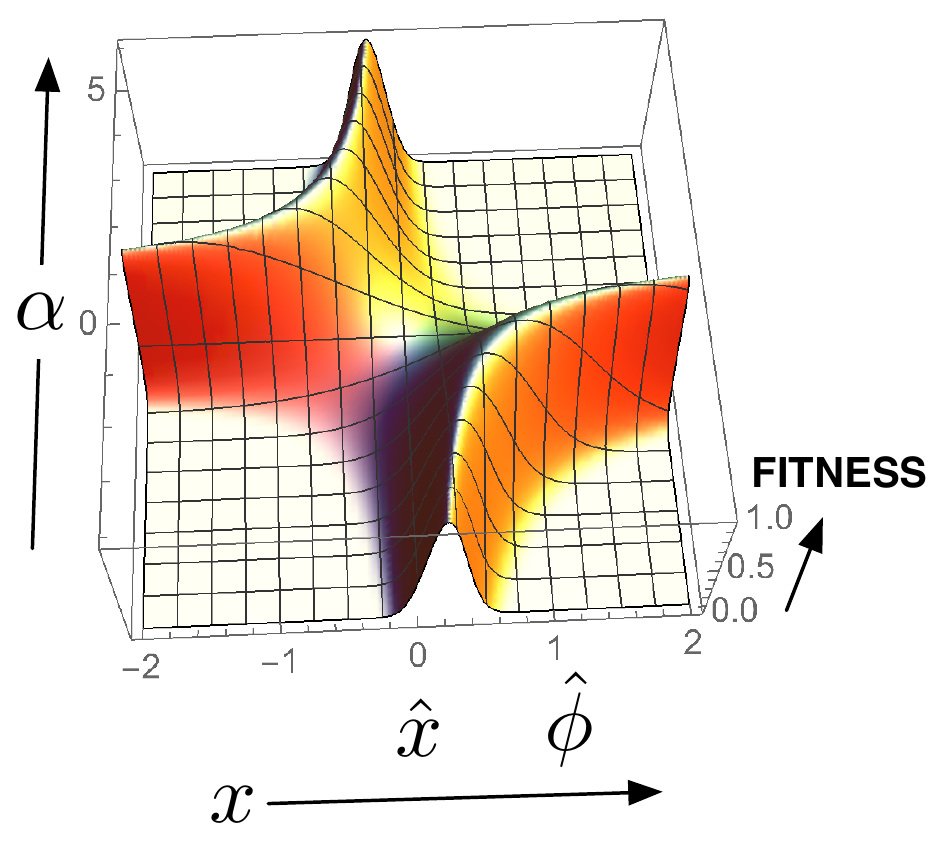}}
\caption{\label{fig:alpha}{\bf Fitness as a function of juvenile phenotype $x \in [-2, 2]$ and plasticity parameter $\alpha \in [-4, 6]$.}  Shown in relation to the population sample $\xh$ set to $\xh=0$, and the optimal phenotype set to $\phih =1$, and $\beta=1$.  This is the univariate phenotype case only. For $\xh < x < \phih$, the optimal values of $\alpha$ to maximize fitness are negative, while for $x < \xh$ or $x > \phih$, the optimal values of $\alpha$ are positive.  The curve for the optimal $\alpha$ is $\alphah = (x-\phih)/(x-\xh)$.}
\end{figure}

Figure \ref{fig:alpha} plots the fitness in the univariate case as a function of juvenile phenotype $x$,  and plasticity parameter $\alpha$ in relation to the population sample $\xh$, and the phenotypic optimum $\phih$.  We see that for $x$ intermediate between $\xh$ and $\phih$, the optimal values of $\alpha$ to maximize fitness are negative (anti-imitation), while for $x$ outside the range $[\xh, \phih]$, the optimal values of $\alpha$ are positive (imitation).  These optimal $\alphah$ produce an adult phenotype $\phi(x) = (1-\alphah) x + \alphah \xh$ which is exactly the optimal phenotype $\phih$.

\subsection{Wright-Fisher Population Dynamics} \label {Wright-Fisher}
A natural choice for a population model of a semelparous organism with discrete, non-overlapping generations is the Wright-Fisher model.  Finite population sampling will be seen to play a crucial role at certain points in the evolutionary dynamics.

We represent the population state as a vector of $N$ random variables that represent each organism, which is a pair $G=(Y, A)$ of its major loci genotype $Y$ and modifier genotype $A$.  For the stabilizing selection model $Y \in \Reals^D$, and in the purifying selection model $Y \in \set{0,1}^L$. To cover both we will say simply that $Y \in \Omega$.  The population vector is $\G \in (\Omega \times \Reals)^N$.  The evolutionary trajectory consists of a sequence $\G(0), \G(1), \G(2), \ldots , \G(t), \ldots$.

\begin{figure}[H]
\centerline{\includegraphics[width=4in]{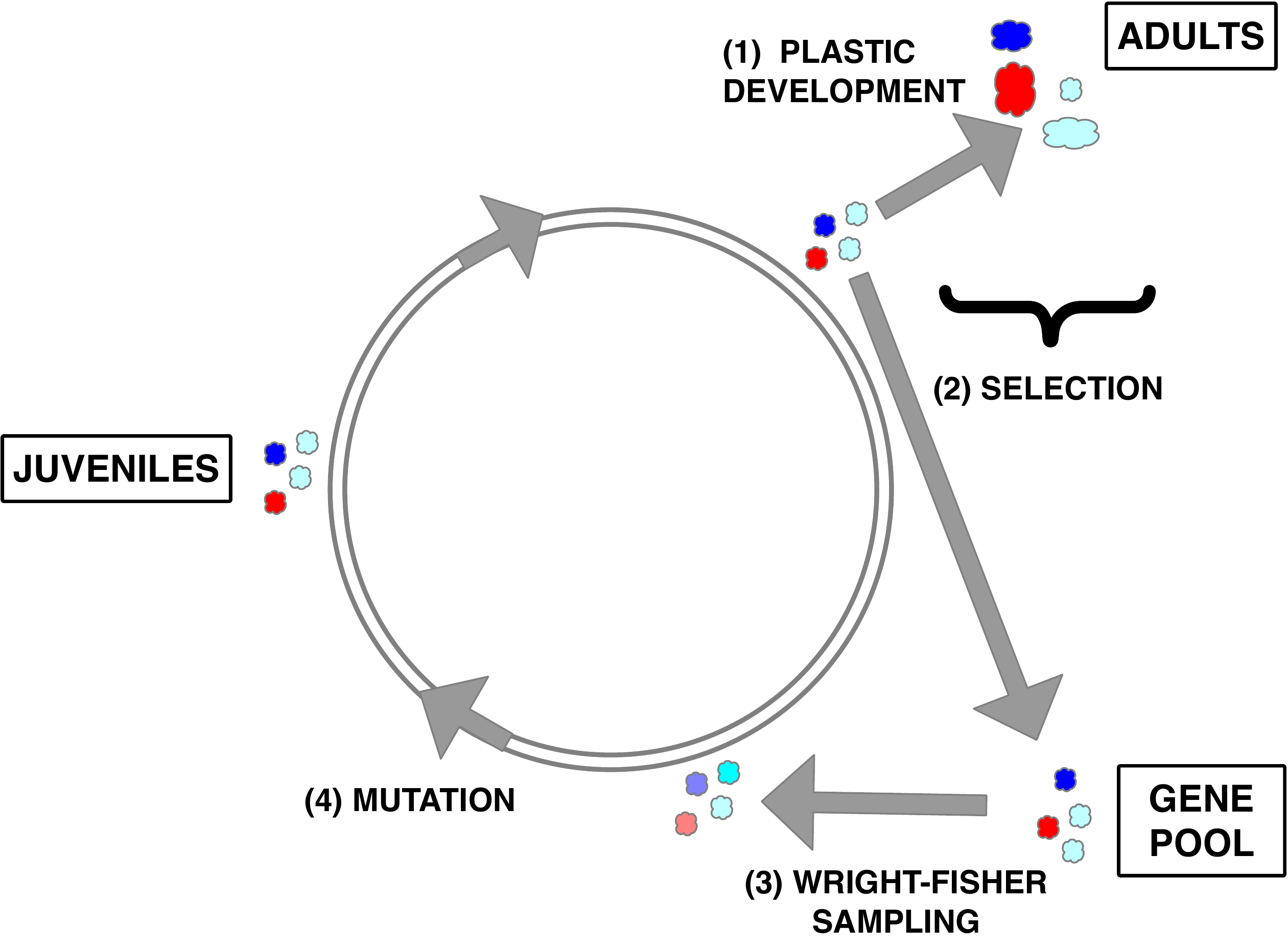}}
\caption{{\bf Depiction of the life cycle.} (1) Juvenile to adult plastic development. (2) Selection on adult phenotype to produce the gene pool. (3) Wright-Fisher sampling. (4) Mutation. \label{fig:LifeCycle}}
\end{figure}

The evolutionary dynamics are captured by the following recurrence relation.  The life cycle is depicted in Figure \ref{fig:LifeCycle}.  After plastic development from juvenile to adult, there is Wright-Fisher sampling of the parent population $\G(t)$ weighted by fitness, to produce a ``gene pool'' $\G^{WF}(t)$:
\an{
G^{WF}_i(t) &= G_j(t), \qquad i, j =1, \ldots, N,& \text{with probability } p_j(t), \label{eq:Wright-Fisher}
}
where each $G^{WF}_i(t)$ is sampled independently from $\G(t)$ with the probability distribution weighted by fitness:
\an{
p_j(t) &= \frac{w(\phi(X_j(t)) }{Z(t)}, && \text{probability proportional to fitness } w(\cdot), \label{eq:Selection}
}
where $X_j(t) = x(Y_j(t))$ is the juvenile phenotype as a function of genotype $Y_j(t)$, and
\ab{
Z(t) &
= \sum_{j=1}^N w (\phi(X_j(t)) ) 
= \sum_{j=1}^N w (\phi(x(Y_j(t)) )), 
&& \text{is the probability normalizer.}
}
The mean fitness of the population is
\an{
\wb(t) &= \frac{1}{N} Z(t) 
= \frac{1}{N} \sum_{j=1}^N w (\phi(X_j(t) ))
= \frac{1}{N} \sum_{j=1}^N w (\phi(x(Y_j(t)) )). 
\label{eq:Wbar}
}
Finally, mutation acts to produce the next generation
\ab{
\G(t{+}1) & = ( G_1(t{+}1), \ldots, G_N(t{+}1) ) \\
&=  ((Y_1(t{+}1), A_1(t{+}1) ), \ldots, (Y_N(t{+}1), A_N(t{+}1) ) )
}
where
\an{
A_i(t{+}1) &= A^{WF}_i(t) + \chi_i , \text{\ \ } \chi_i \sim \Normal(0, \sigma^2) \text{ (Gaussian mutation)}, \label{eq:MutationA}\\
\ \\
Y_i(t{+}1) &= \!
\Cases{\!\!\!
Y^{WF}_i(t) + \xi_i,  \text{\ } \xi_i \sim \Normal(\0, \sigma^2 \I) \text{ (stabilizing selection model)}, \\
\ \\
\!\!\!(Y^{WF}_i(t) + \beta_i) \moddd 2,  \text{ (purifying selection model)}
} \label{eq:MutationY}
}
with $\beta_i \in \set{0,1}^L$, and each $\beta_{ij}$, $j=1, \ldots, L$, is sampled independently with probabilities:
\ab{
\Prob[\beta_{ij} = 1] &= \mu \\
\Prob[\beta_{ij} = 0] &= 1-\mu.
}
Modular arithmetic is employed as a simple way to implement mutations using addition.

The mutation rates used here are large enough to keep the populations polymorphic in the face of strong natural selection.  These models would thus be classified as in the ``strong mutation, strong selection'' regime.  The alternative ``weak mutation, strong selection'' regime not studied here would leave the population monomorphic for most generations, experiencing rapid selective sweeps of new advantageous mutations.  

\section{Results}

We organize the results into two main sections:  \ref{sub:Fixed} {\bf Speed of Adaptation and Genetic Load Under Fixed Imitation} examines evolution of the juvenile phenotypes under different fixed values of the imitation parameter $\alpha$; and \ref{sub:Evolves} 
{\bf Evolution of Imitation and Anti-Imitation}  examines joint evolution of $\alpha$ and the juvenile phenotypes.  Within each section we examine the {\bf stabilizing selection model}
 \ref{subsub:FixedStabilizing} \ref{subsub:EvolvesStabilizing}
and the {\bf purifying selection model}.
 \ref{subsub:FixedPurifying} \ref{subsub:EvolvesPurifying}.

\subsection{Speed of Adaptation and Genetic Load Under Fixed Imitation}\label{sub:Fixed}

Here we examine the evolutionary trajectories of the juvenile phenotype under different fixed values of the imitation parameter $\alpha$.

\subsubsection{Stabilizing Selection Model}\label{subsub:FixedStabilizing}
\begin{figure}[H]
\begin{subfigure}{0.5\textwidth}
\centerline{\includegraphics[width=\linewidth]{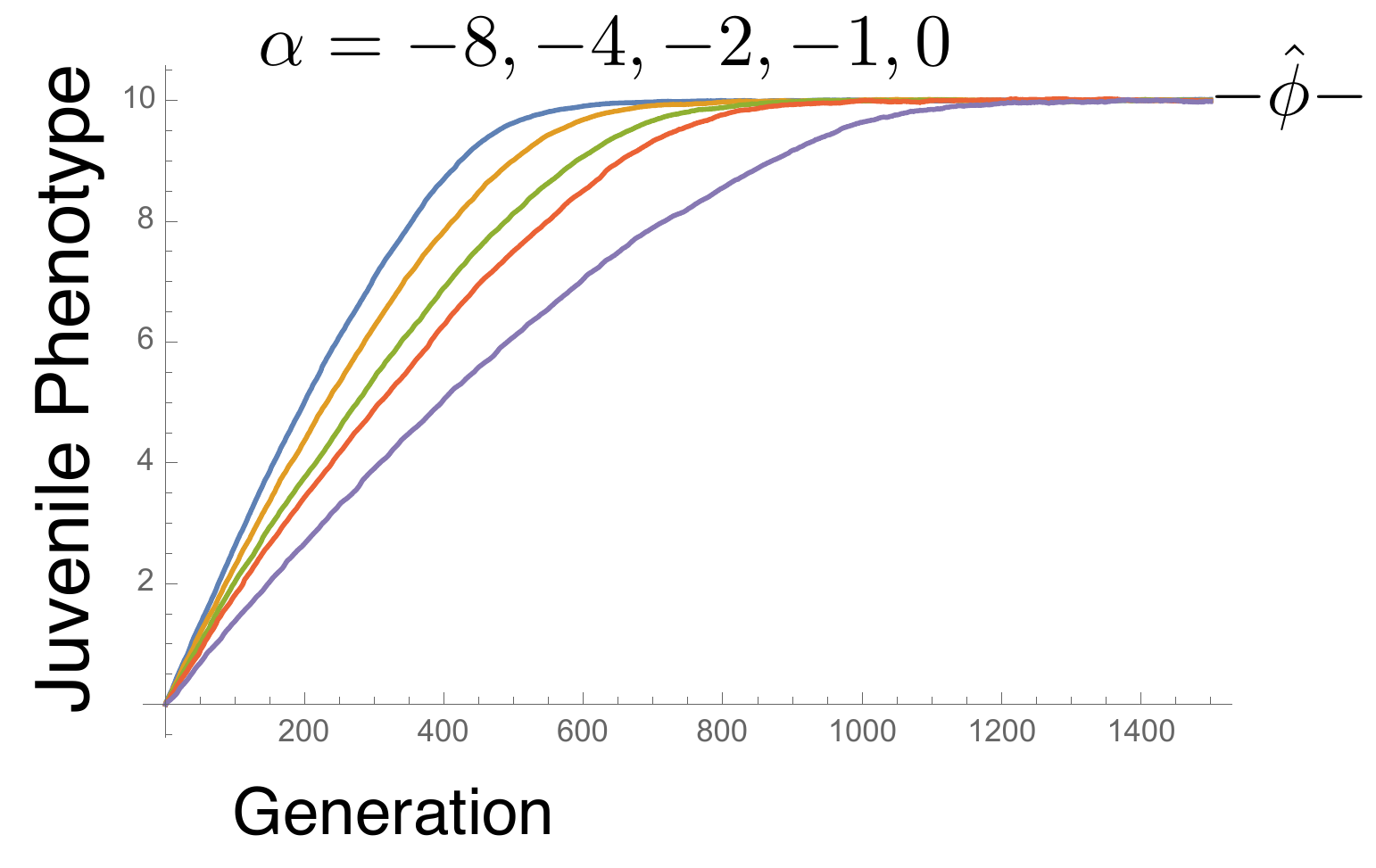}}
\end{subfigure}
\begin{subfigure}{0.5\textwidth}
\centerline{\includegraphics[width=\linewidth]{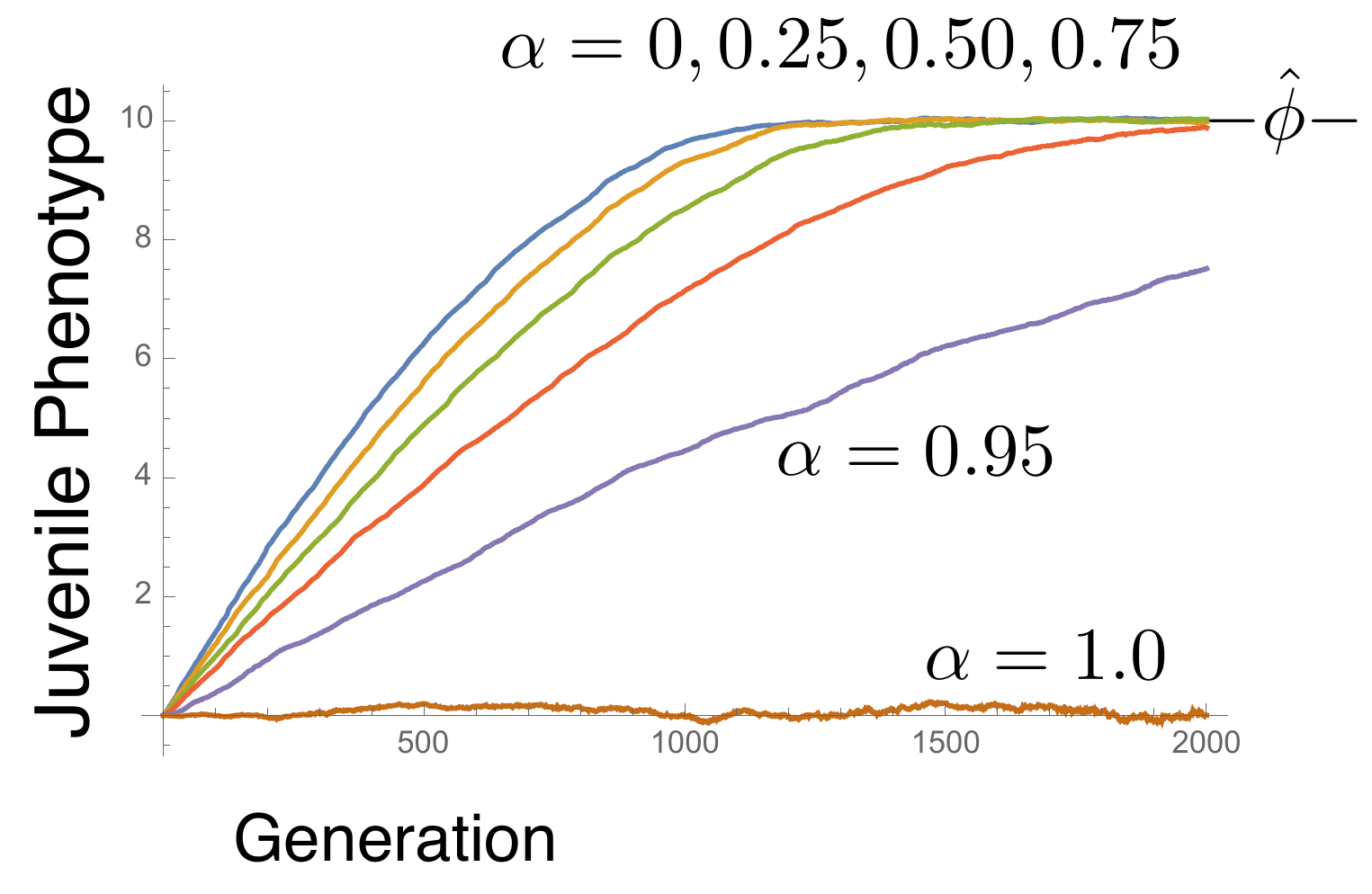}}
\end{subfigure}
\caption{{\bf Evolutionary trajectories of the population mean juvenile phenotype.} For four negative values of $\alpha$ (left), and five positive values of $\alpha$ (right), with $\alpha=0$ in both.  Other parameters: $\phih = 10$, $N=1024$, $\beta=1$, $\sigma^2= 0.01$, $D=1$.\label{fig:FixedAlphaPhene}}
\end{figure}

First we investigate the evolutionary trajectories for fixed imitation parameters, where there is no evolution in $\alpha$.  We initialize the juvenile phenotypes far from the phenotypic optimum, so that they begin in a state of directional selection.
Figure \ref{fig:FixedAlphaPhene} shows evolutionary trajectories for five values of $\alpha$ in the negative range (left graph) and in the positive range (right graph). We see that the more negative $\alpha$ is---i.e. the greater the anti-imitation---the faster is the evolution of the juvenile phenotypic toward the optimal phenotype.  In the left graph we see the transition in each trajectory from directional selection to stabilizing selection, arrived at sooner for more negative $\alpha$.   In the right graph, we see the continual slowing down of phenotypic evolution as $\alpha$ increases in the positive range $\alpha \in \{0, 0.25, 0.5, 0.75, 0.95\}$. Finally, at $\alpha=1.0$, there is genotype-phenotype disengagement, and the juvenile phenotype evolves strictly under genetic drift.

Why does anti-imitation produce faster evolution under directional selection?  Anti-imitation repels plastic phenotypes away from the population mean.  So juvenile phenotypes worse than the population mean produce adult phenotypes that are worse still, while juvenile phenotypes that are better than the population mean produce adult phenotypes that are better still.  Anti-imitation therefore disadvantages disadvantageous genotypes, and advantages advantageous genotypes---the ``Matthew effect'' \citep{Merton:1968:Matthew}.  Anti-imitation thus amplifies the strength of directional selection.

More insight comes from the effect on the distribution of fitness effects of mutation.  Organisms with the highest probability of generating mutant offspring fitter than the mean $\wb$ are also those closer to the optimum $\phih$ than is the population mean $\xb$.  These organisms are the likely ancestors of the eventual future population, and it is these organisms whose fitnesses are increased by anti-imitation.  Meanwhile, organisms with below-average fitness---further from $\phih$ than is $\xb$---are much less likely to give rise by mutation to the descendent population, and it is these organisms whose fitnesses are decreased by anti-imitation.  Therefore, anti-imitation increases evolvability \citep{Altenberg:1994:EEGP,Altenberg:1995:GGEGPM} by increasing selection for individuals most likely to give rise to the fittest in the next generation.

\begin{figure}[H]
\centerline{\includegraphics[width=3.6in]{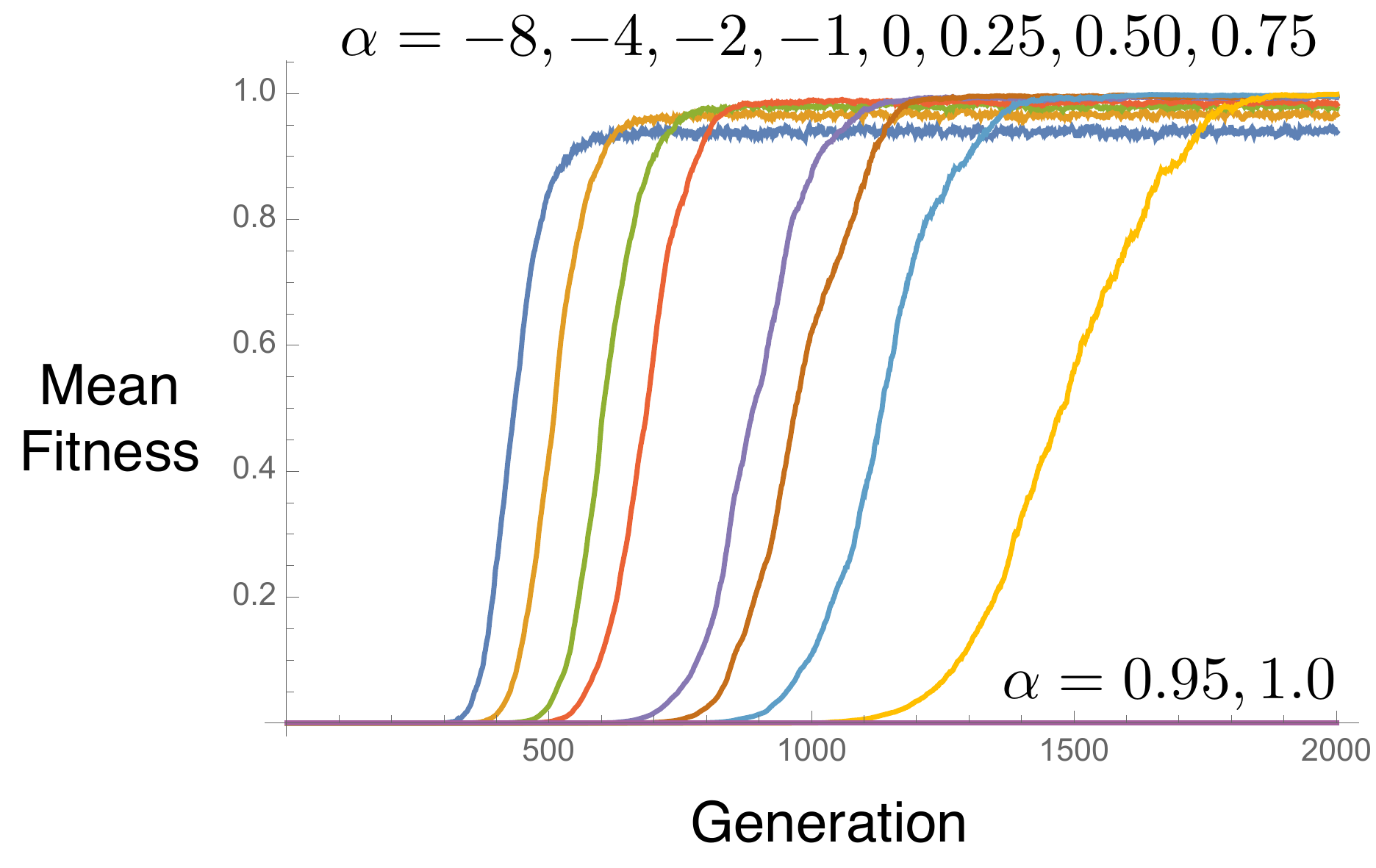}}
\caption{{\bf Evolutionary trajectories in the population mean fitness $\wb(t)$.}  For $\alpha \in \set{ -8, -4, -2, -1, 0, 0.25, 0.5, 0.75, 0.95, 1.0}$.  Other parameters: $\phih = 10$, $N=1024$, $\beta=1$, $\sigma^2= 0.01$, $D=1$.\label{fig:FixedAlphaWbar}} 
\end{figure}

Figure \ref{fig:FixedAlphaWbar} shows the evolution of the population mean fitness for the whole range $\alpha \in \{ -8, -4$, $-2, -1, 0, 0.25, 0.5, 0.75, 1.0\}$.  Here we see that more anti-imitation produces the most rapid increases in mean fitness.  For $\alpha=1.0$, which produces genotype-phenotype disengagement, there is no increase in mean fitness.  For $\alpha=0.95$, mean fitness increases, but because it is so slow to evolve toward the phenotypic optimum, it is not visibly different from $0$ in the plot.  
An important new phenomenon appears in the stabilizing selection phase: a tradeoff between the speed of adaptation and the equilibrium \emph{genetic load} (suppression of mean fitness below the maximum possible fitness).  The value  $\alpha = -8$ produces the fastest increase in mean fitness during directional selection, but it also produces the greatest genetic load during stabilizing selection.  The fluctuations in mean fitness seen during the stabilizing selection phase in Figure \ref {fig:FixedAlphaWbar} are due to finite population sampling effects.

\begin{figure}[H]
\centerline{\includegraphics[width=3in]{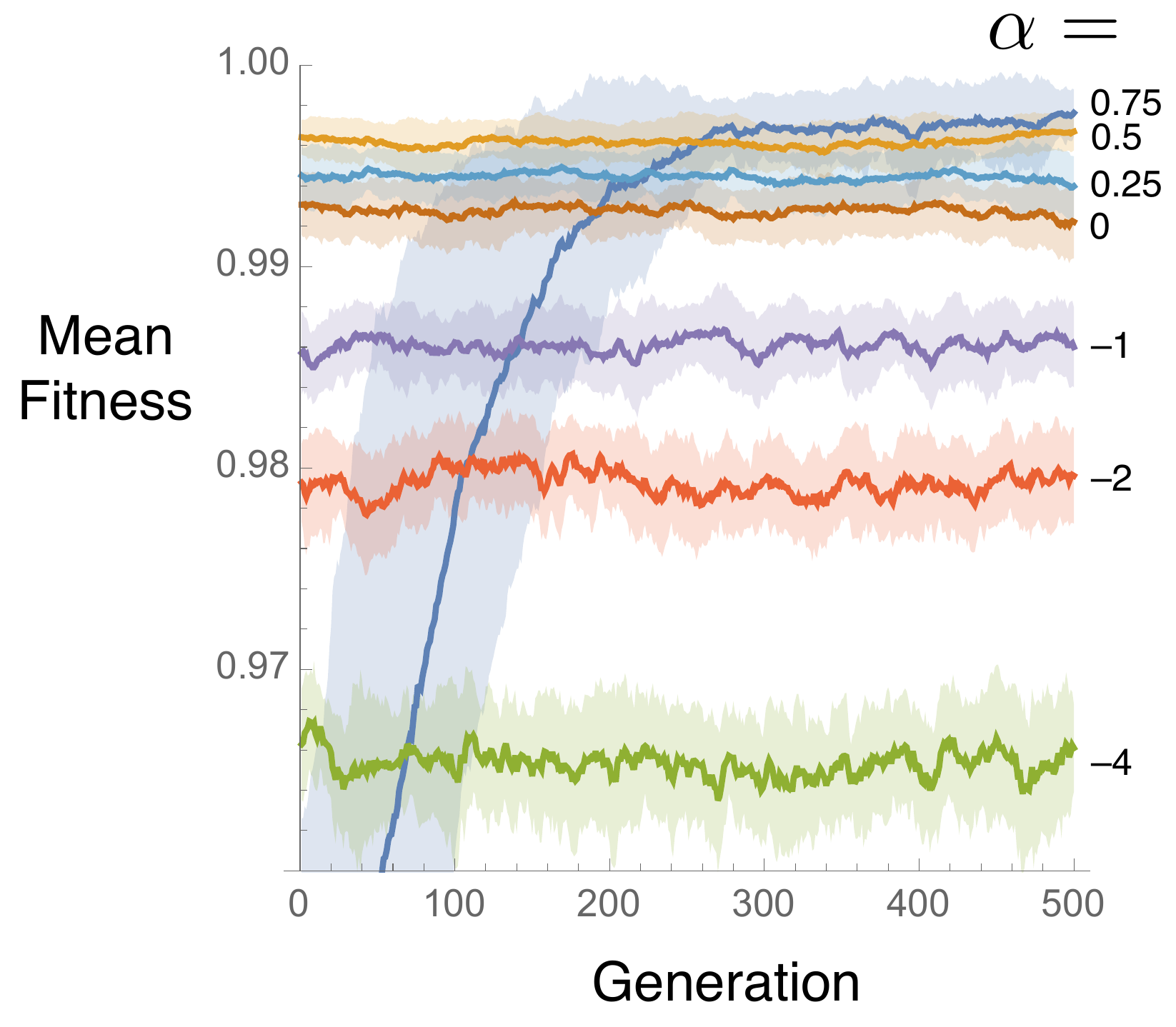}}
\caption{{\bf Zoom in on Fig.\  \ref {fig:FixedAlphaWbar}.}  Showing the population mean fitness $\wb(t)$ averaged over 24 evolutionary trajectories for each $\alpha \in \set{  -4, -2, -1, 0, 0.25, 0.5, 0.75}$.  Shaded bands are $\pm$ one standard deviation. Shown are generations 1700--2200 continuing Fig.\  \ref {fig:FixedAlphaWbar} (labeled 0--500).  Other parameters: $\phih = 10$, $N=1024$, $\beta=1$, $\sigma^2= 0.01$, $D=1$.\label{fig:FixedAlphaLoad}}
\end{figure}
The amount of genetic load for different values of $\alpha$ is shown in Figure \ref{fig:FixedAlphaLoad}.  The averages of 24 repeated evolutionary trajectories are shown for each value of $\alpha$.  The figure corresponds to a closeup of Figure \ref{fig:FixedAlphaWbar} for the last 300 generations of the evolutionary trajectories, and extends it by 200 generations.  We see that  in this stabilizing selection phase, the mean fitnesses of the populations increase with $\alpha$.  However, for $\alpha = 0.75$ the speed of evolution is so slow that it reaches stabilizing selection after all the other trajectories;  it crosses the trajectories of the smaller values of $\alpha$ to emerge as the highest mean fitness on the right.  Due to stochasticity from finite sampling, the standard deviations of the trajectories for $\alpha = 0, 0.25, 0.5$, and $0.75$ considerably overlap. 
\begin{figure}[H]
\centerline{\includegraphics[width=4in]{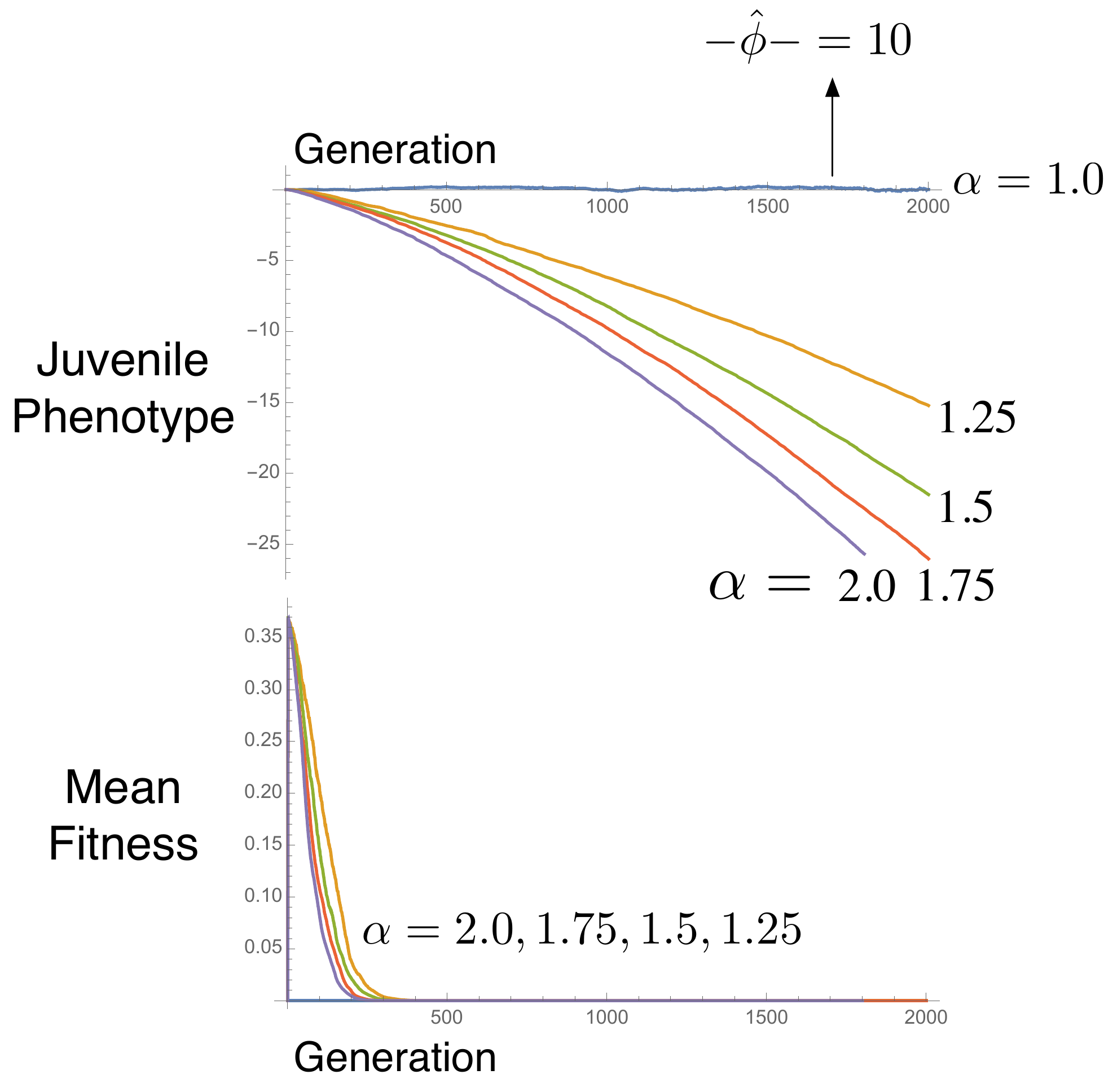}}
\caption{{\bf Evolutionary trajectories of the population mean juvenile phenotype for values of $\alpha$ in the hyper-imitation range beyond 1.0.}  The plasticity function with $\alpha > 1$ causes the adult phenotype $\phi$ to overshoot the population sample phenotype $\xb$. Thus juvenile phenotypes $x$ farthest below $\xb$ have adult phenotypes closest to $\phih=10$, making them the fittest.  Evolution pathologically sends juvenile phenotypes racing away from the optimum (top graph) which causes the population mean fitness to plummet in time (bottom graph), faster with greater $\alpha$.  Other parameters: $\phih = 10$, $N=1024$, $\beta=1$, $\sigma^2= 0.01$, $D=1$.\label{fig:FixedAlphaPheneHyper}}
\end{figure}

So we see a tradeoff between rate of evolution under directional selection and genetic load under stabilizing selection. 

Figure \ref {fig:FixedAlphaPheneHyper} shows the evolution of juvenile phenotypes in the hyper-imitation range $\alpha \in \{1, 1.25$, $1.5, 1.75, 2.0\}$.  Here evolution takes the juvenile phenotypes in the opposite direction from the optimal phenotype $\phih =10$, a pathological situation in which stabilizing selection never arrives.
For the pathological cases of $\alpha \in \set{1.25, 1.5, 1.75, 2.0}$ shown in Figure \ref {fig:FixedAlphaPheneHyper}, the mean fitness of the population actually evolves to \emph{decrease}.  This can be interpreted as an effect of frequency-dependent selection, because the adult phenotypes are frequency-dependent. It is well-known that under frequency-dependent selection, the mean fitness need not increase under evolution \citep{Turner:1967:Supergenes,Altenberg:1991,Asmussen:etal:2004:FDS}.  The steady decrease in mean fitness as a response to natural selection can be considered to be an ``evolutionary pathology'' \citep{Nunney:1999:Lineage:benefit,Altenberg:2005:Evolvability}.  

So we have a situation where the long-term mean fitness increases as $\alpha$ approaches $1$, but when $\alpha$ exceeds $1$, evolution pushes the population away from the optimum in this pathological manner. 

\subsubsection{Purifying Selection Model}\label{subsub:FixedPurifying}

In this section we investigate the model for purifying selection, where the juvenile phenotype is determined by a multilocus bi-alleleic (0,1) genotype in which fitness increases as a function of the number of $1$ alleles in the genotype.  The juvenile phenotype $x$ is the fraction of alleles in state $1$.  The number of loci is set to $L=128$.  

Selection is for the extreme phenotype $\phih=1.0$, which only the all-1s genotype can achieve, all other genotypes producing juvenile phenotype of lesser value.  Thanks to the phenotypic plasticity function, however, adult phenotypes have the potential to reach $1.0$.  Fitness is again a Gaussian function of the departure of the adult phenotype from $\phih = 1.0$.  Since the juvenile phenotype falls in the small range $x \in [0, 1]$, we rescale the selection intensity to $\beta=10$ to make it roughly equivalent to the stabilizing selection model.

\begin{figure}[H]
\centerline{\includegraphics[width=5in]{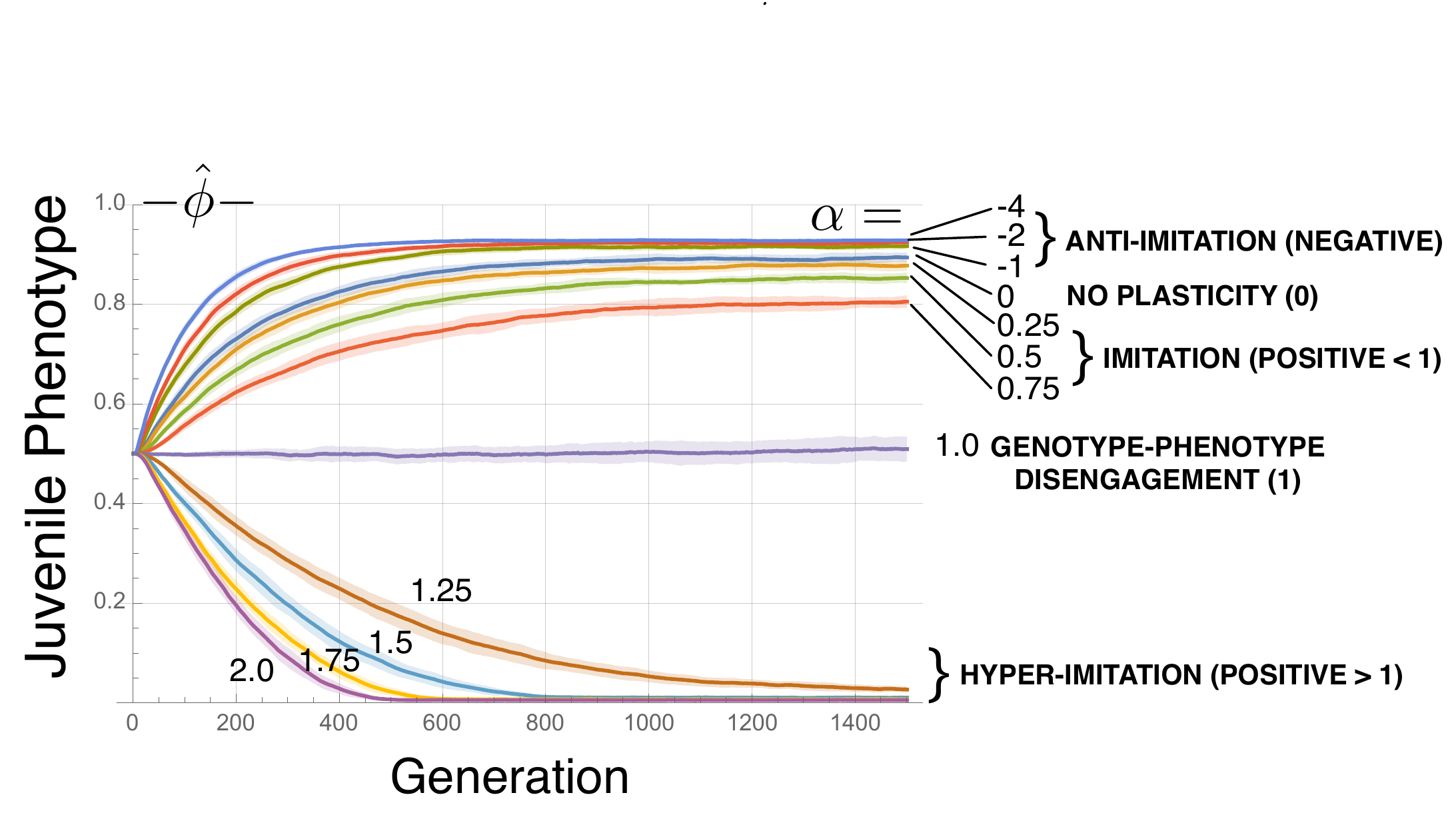}}
\caption{{\bf Evolutionary trajectories in the mean population juvenile phenotype $\xb(t)$ for the multilocus purifying selection model.}  Plotted are the average of 24 runs.  The shaded areas are $\pm$ one standard deviation.  The trajectories are labeled by $\alpha$ and the range they belong to: anti-imitation, imitation, and hyper-imitation.  Other parameters: $\phih = 1.0$, $N=1024$, $\beta=10$, $\mu= 0.1/L$, $L=128$.  \label{fig:FixedAlphaPurPhene}}
\end{figure}

Populations are initialized with genotypes that are half $0$ alleles and half $1$ alleles, which is the equilibrium genotype under mutation alone.  Figure \ref{fig:FixedAlphaPurPhene} shows the evolution of the juvenile phenotypes for the range of $\alpha \in \{-4, -2, -1, 0, 0.25$, $0.5, 0.75, 1.0, 1.25, 1.5, 1.75, 2.0\}$.  Plotted are the averages of 24 independent runs for each parameter value.  Shaded bands are $\pm$ one standard deviation.

We see that the evolutionary trajectories are ordered by their $\alpha$ value, and maintain their order throughout the generations.  At each time point the trajectories are monotonically decreasing in $\alpha$. 

Strong anti-imitation, $\alpha < 0$, yields the fastest evolution and phenotypes closest to $\phih=1$ at mutation-selection balance.  Genotype-phenotype disengagement is produced by $\alpha = 1$, and we see the genotypes remain near the pure-mutation balance of half $0$s, half $1$s, to give juvenile phenotypes near $0.5$.  Hyper-imitation, $\alpha > 1$, again produces an evolutionary pathology: the juvenile phenotypes evolve \emph{away} from the optimum $\phih =1.0$ toward $0$, faster with greater hyper-imitation.

\begin{figure}[H]
\centerline{\includegraphics[width=5in]{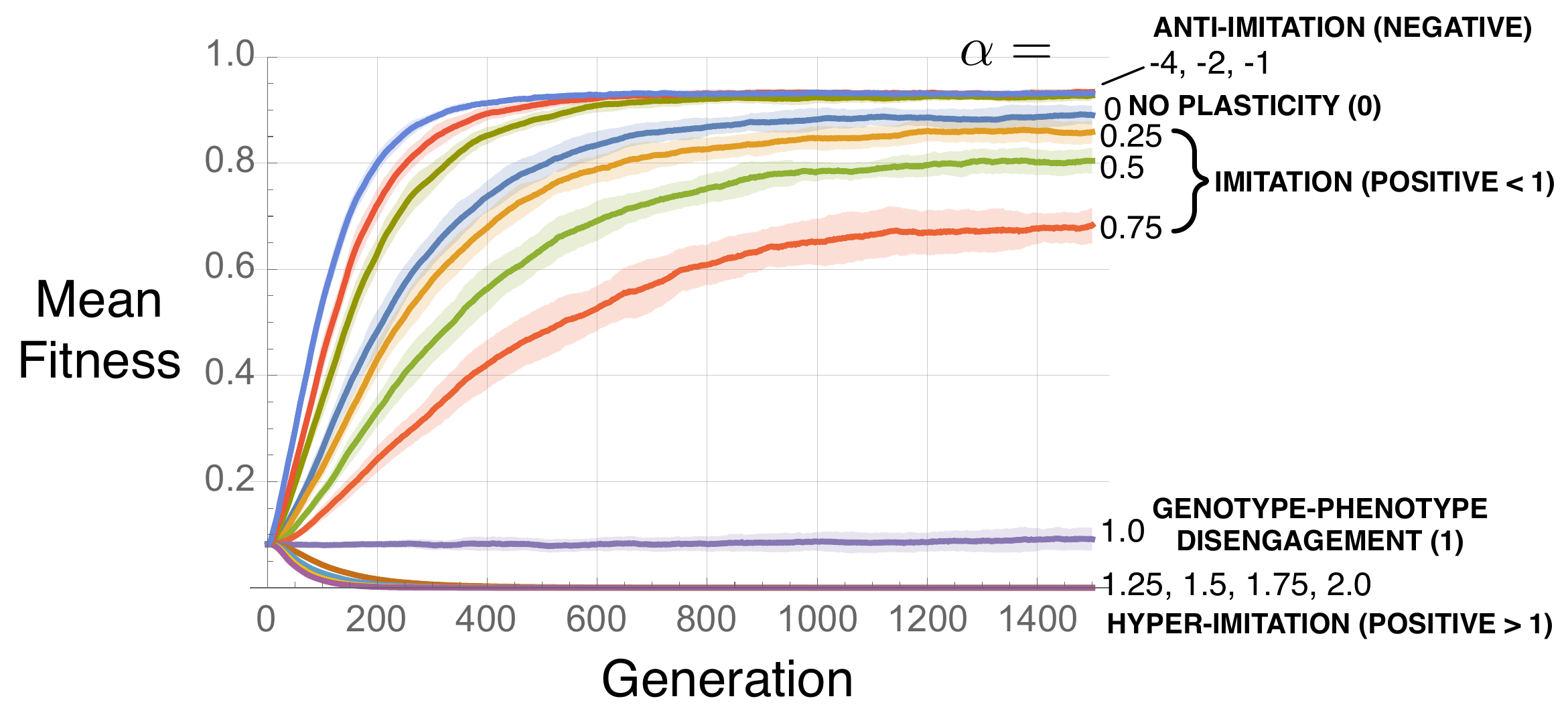}}
\caption{{\bf Evolutionary trajectories in the population mean fitness $\wb(t)$ for the multilocus purifying selection model.}   For $\alpha$ ranges over (top to bottom) $\{ -8, -4, -2, -1, 0, 0.25, 0.5, 0.75, 0.95, 1.0, 1.25, 1.5, 1.75, 2.0\}$.  Other parameters: $\phih = 10$, $N=1024$, $\beta=10$, $\mu= 0.1/L$, $L=128$. No tradeoff is observed between speed of evolution and equilibrium genetic load.  Rather, both decrease monotonically with $\alpha$. \label{fig:FixedAlphaPurWbar}}
\end{figure}

When we examine the population mean fitnesses for the purifying selection model, we see a very different outcome from the stabilizing selection model.  Figure \ref{fig:FixedAlphaPurWbar} shows the mean fitness $\wb(t)$ for the range of $\alpha$ from the average of 24 evolutionary trajectories.  We no longer see the tradeoff between speed of evolution and equilibrium mean fitness that we observed with stabilizing selection (Figures \ref{fig:FixedAlphaWbar}, \ref{fig:FixedAlphaLoad}).  Figure \ref{fig:FixedAlphaPurWbar} shows that anti-imitation gives not only the fastest adaptation, but also the highest mean fitness at mutation-selection balance.  As expected from the graphs of the phenotypes, the highest mean fitnesses at mutation-selection balance are attained with anti-imitation, and the genetic load increases with increasing $\alpha$.  The genetic load increases dramatically as $\alpha$ approaches $1$ which gives genotype-phenotype disengagement.  For all the values of $\alpha > 1$ in the hyper-imitation range, the mean fitness at mutation-selection balance is near zero.

This can be understood because with purifying selection for an extreme phenotype, the juvenile phenotypes $x$ are always less than the optimum $\phih=1$.  The population is therefore always in a state similar to directional selection in that the juvenile phenotypes are all on one side of the optimum.  The juvenile phenotypes that are above average ($x > \xb$) have their adult phenotypes brought closer to the optimum $\phih=1$ by anti-imitation, whereas imitation shifts their adult phenotypes away from the optimum.  The reverse holds for juvenile phenotypes below average ($x < \xb$).  In  Section \ref{subsub:EvolvesPurifying} we will see that evolution favors the modifier allele which improves the fitness of the best genotypes even while it lessens the fitnesses of the worst genotypes.
 
\subsection{Evolution of Imitation and Anti-Imitation}\label{sub:Evolves}

In the previous section, we examined how populations evolve under different fixed values of the imitation parameter $\alpha$.  
In this section we follow the ``strategy of endogenization'' \citep{Okasha:2018:Strategy} and introduce genetic variation in the modifier locus controlling $\alpha$ \citep{Fisher:1931:Modification,Nei:1967:Modification,Feldman:1972}.  We ask how the $\alpha$ values evolve under the same population dynamics as before except that now the modifier locus controlling $\alpha$ is subject to mutation and is no longer fixed.

We will see that, for the most part, those values of $\alpha$ that gave rise to greater mean fitnesses are the same values that give a selective advantage to genetic variation of the modifier gene.  However, as very large value of imitation evolve we encounter paradoxical and complex evolutionary dynamics that are not derivable from mean fitness arguments, showing the importance of specific mechanistic modeling for the evolution of non-fitness parameters.

\subsubsection{Stabilizing Selection Model}\label{subsub:EvolvesStabilizing}
\begin{figure}[H]
\centerline{\includegraphics[width=4in]{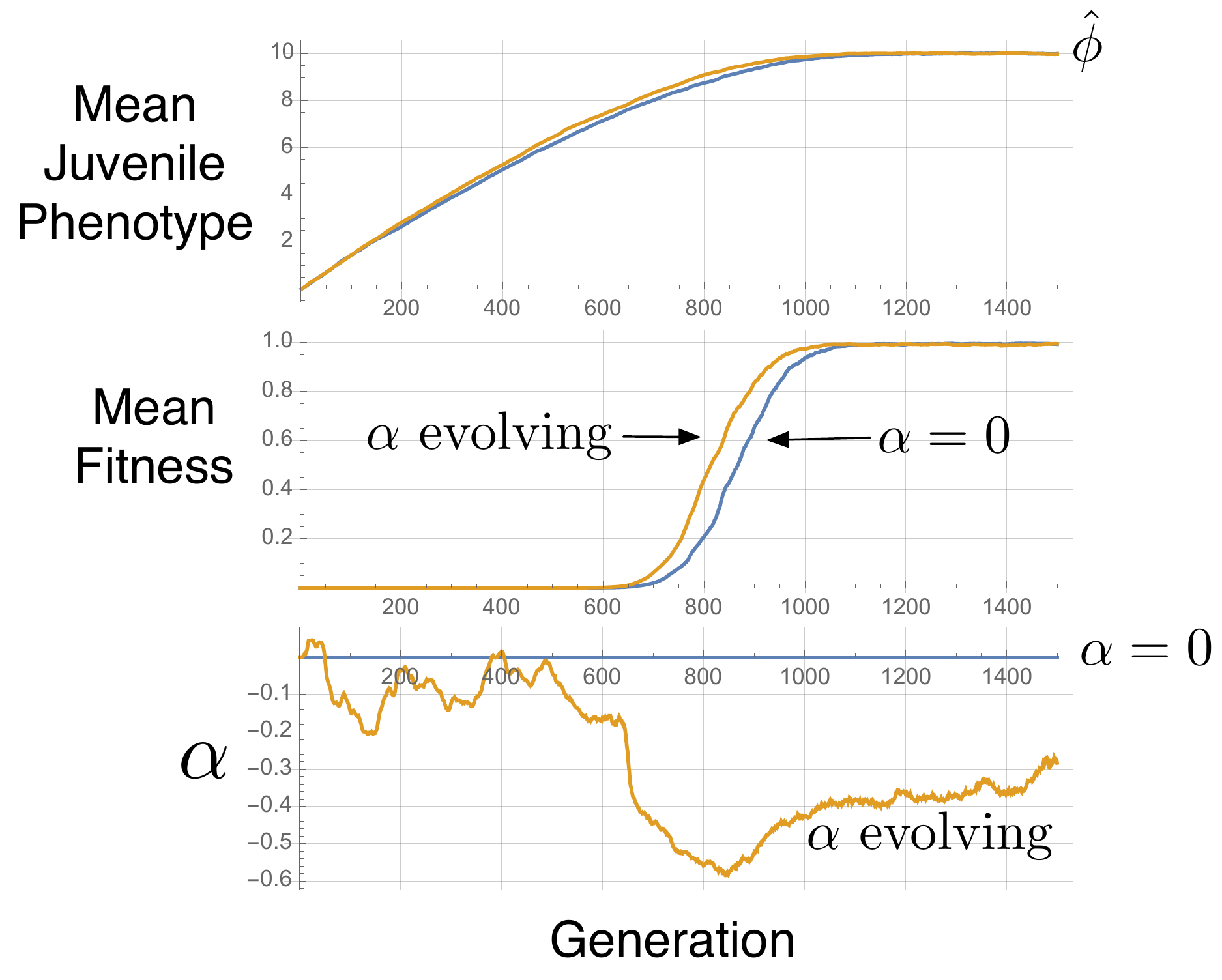}}
\caption{{\bf Evolutionary trajectories for 1500 generations.}  Showing the population mean phenotype $\xb(t)$, fitness $\wb(t)$, and $\alpha$, in orange.  For comparison, an evolutionary run with fixed $\alpha=0$ is plotted in blue.  Other parameters: $\phih = 10$, $N=1024$, $\sigma^2= 0.01$, $D=1$.\label{fig:AlphaEvolves}}
\end{figure}

Figure \ref {fig:AlphaEvolves} shows the evolution of $\alpha$ during a single evolutionary run of the stabilizing selection model, in which the population begins far from the optimal phenotype and experiences a period of directional selection, but finally reaches a mutation-selection balance surrounding the phenotypic optimum.  The population is initialized at $\alpha=0$ and $x = 0$ with a phenotypic optimum of $\phih =10$.  Both $\alpha$ and $x$ mutate by addition of a Gaussian random variable with mean $0$ and variance $\sigma^2 = 0.01$.  Shown are the average population juvenile phenotype $\xb(t)$, the mean fitness $\wb(t)$, and the mean $\alpha$ (in orange). For comparison, the trajectories for fixed $\alpha=0$ are plotted in blue.

The rate of evolution in the lineage with evolving $\alpha$ is sped up by around 50 generations compared to the lineage with fixed $\alpha=0$.  We see that $\alpha$ itself evolves along a rather meandering course, but goes clearly negative (anti-imitation) until directional selection ceases and stabilizing selection begins, at which point $\alpha$ begins to increase.  This is consistent with previous results for evolution under different fixed values of $\alpha$.

\begin{figure}[H]
\centerline{\includegraphics[width=3.6in]{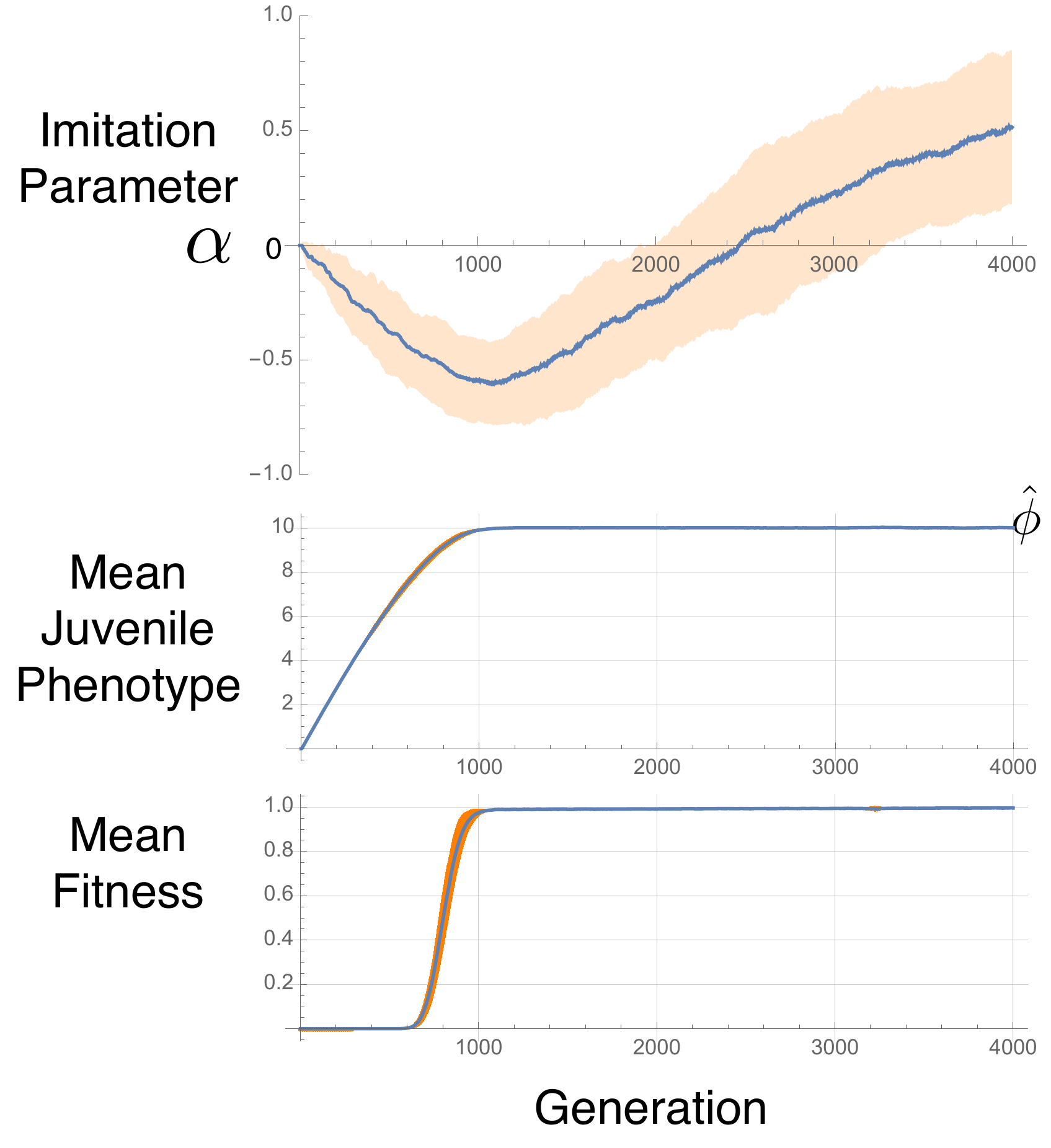}}
\caption {The average of 24 evolutionary trajectories for 1500 generations showing the evolution of the population mean $\alpha$.  Corresponding evolution of the mean juvenile phenotype $\xb(t)$, and mean  fitness $\wb(t)$ are shown.  Orange bands are $\pm$ one standard deviation. Anti-imitation $\alpha < 0$ evolves during the period of directional selection (generations 0 to $\approx 1000$).  When the juvenile phenotype has reached the phenotypic optimum $\phih =10$, stabilizing selection ensues, and $\alpha$ evolves to steadily increase from anti-imitation to imitation ($\alpha > 0$). Other parameters: $N=1024$, $\sigma^2= 0.01$, $D=1$.\label{fig:AlphaEvolvesAvg}} 
\end{figure}

\begin{figure}[H]
\centerline{\includegraphics[width=4in]{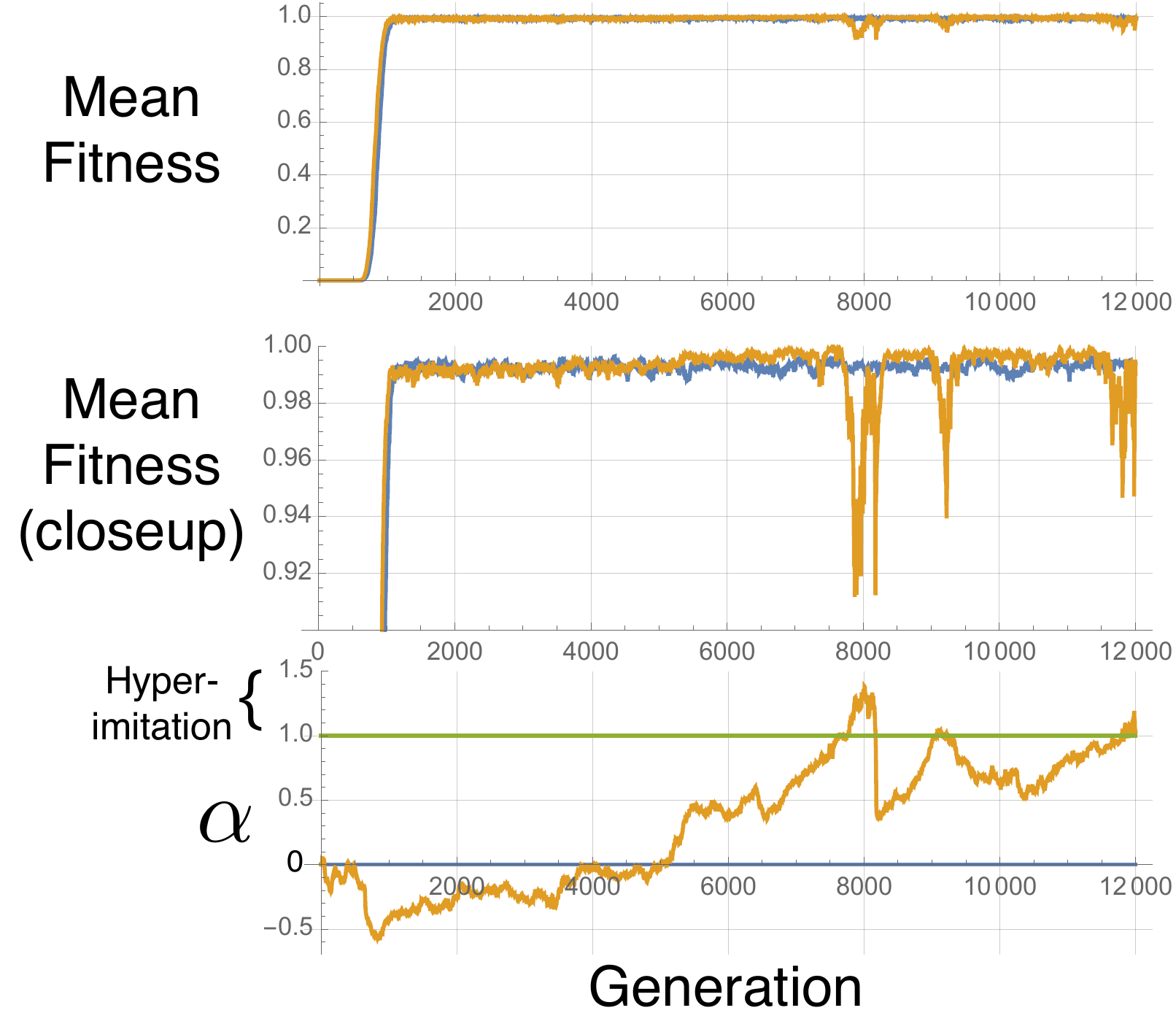}}
\caption{{\bf Extended evolutionary trajectories to 12,000 generations.}  Showing the population mean  fitness $\wb(t)$ (top), zoomed in (middle), and mean evolved $\alpha$ (bottom), in orange.  For comparison, an evolutionary run with fixed $\alpha=0$ is plotted in blue behind the orange. During directional selection $\alpha$ evolves negative. During stabilizing selection, $\alpha$ evolves to increase, with a great deal of fluctuation. When $\alpha$ evolves into the hyper-imitation range, the mean fitness drops, and a rapid collapse of $\alpha$ ensues.  Other parameters: $\phih = 10$, $N=1024$, $\sigma^2= 0.01$, $D=1$.\label{fig:AlphaEvolvesZoom}}
\end{figure}

To get a better idea of the trends in $\alpha$ evolution, we show the average $\alpha$ values from 24 runs of the evolutionary trajectories in Figure \ref {fig:AlphaEvolvesAvg}.  Now it can be clearly seen that $\alpha$ evolves to decrease for the duration of directional selection, and as soon as the juvenile phenotypes reach the phenotypic optimum $\phih =10$ and stabilizing selection takes hold, $\alpha$ evolves to increase until the end of the simulation.  
To see what happens after long periods of stabilizing selection, the evolutionary trajectories are extended to 12,000 generations in Figure \ref {fig:AlphaEvolvesZoom}.  There we see the initial evolution of $\alpha$ into anti-imitation during directional selection, followed by upward evolution in a meandering fashion throughout the period of stabilizing selection, into the imitation range $\alpha > 0$, until it reaches genotype-phenotype disengagement ($\alpha = 1$), but then goes further into hyper-imitation ($\alpha > 1$).  At this point, around generation 7,700, something unusual happens.  We see in the top graph that there is a sudden drop in mean fitness.  This is shown in detail in the middle graph where the $Y$ axis is enlarged. There is a steep plunge in the mean fitness when $\alpha$ enters the hyper-imitation zone.
This is followed at around generation 8,100 by a dramatic collapse in $\alpha$ from it peak of around 1.3, down to 0.4 in the middle of the imitation range.  This collapse in $\alpha$ results in a sudden restoration of the mean fitness to its maximum range, greater than the equilibrium mean fitness seen with fixed $\alpha=0$.  After this point, $\alpha$ is again seen to increase until it reaches hyper-imitation, upon which the mean fitness shows a sudden decline, and $\alpha$ begins a more gradual decline into the imitation range, followed by another rise to hyper-imitation.

\begin{figure}[H]
\begin{subfigure}{0.5\textwidth}
\centerline{\includegraphics[width= \linewidth]{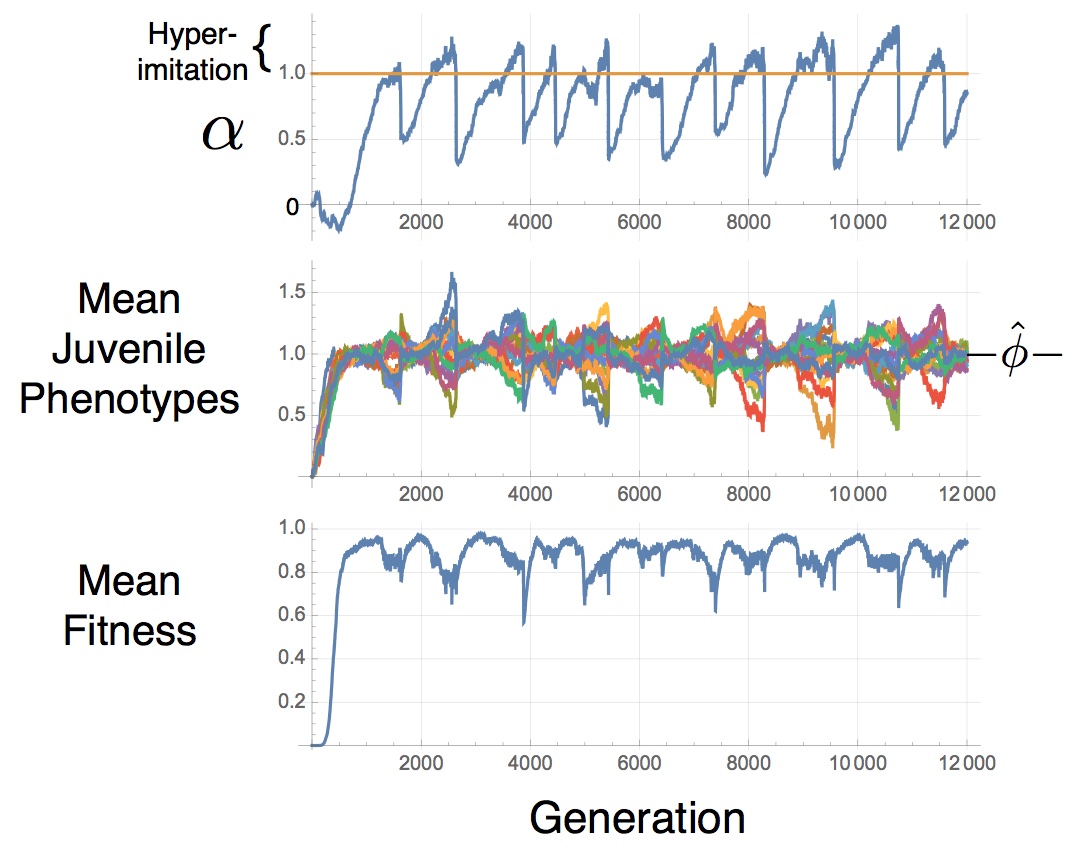}} 
\subcaption{$\alpha$ free\label{fig:AlphaEvolves16}}
\end{subfigure}
\begin{subfigure}{0.5\textwidth}
\centerline{\includegraphics[width= \linewidth]{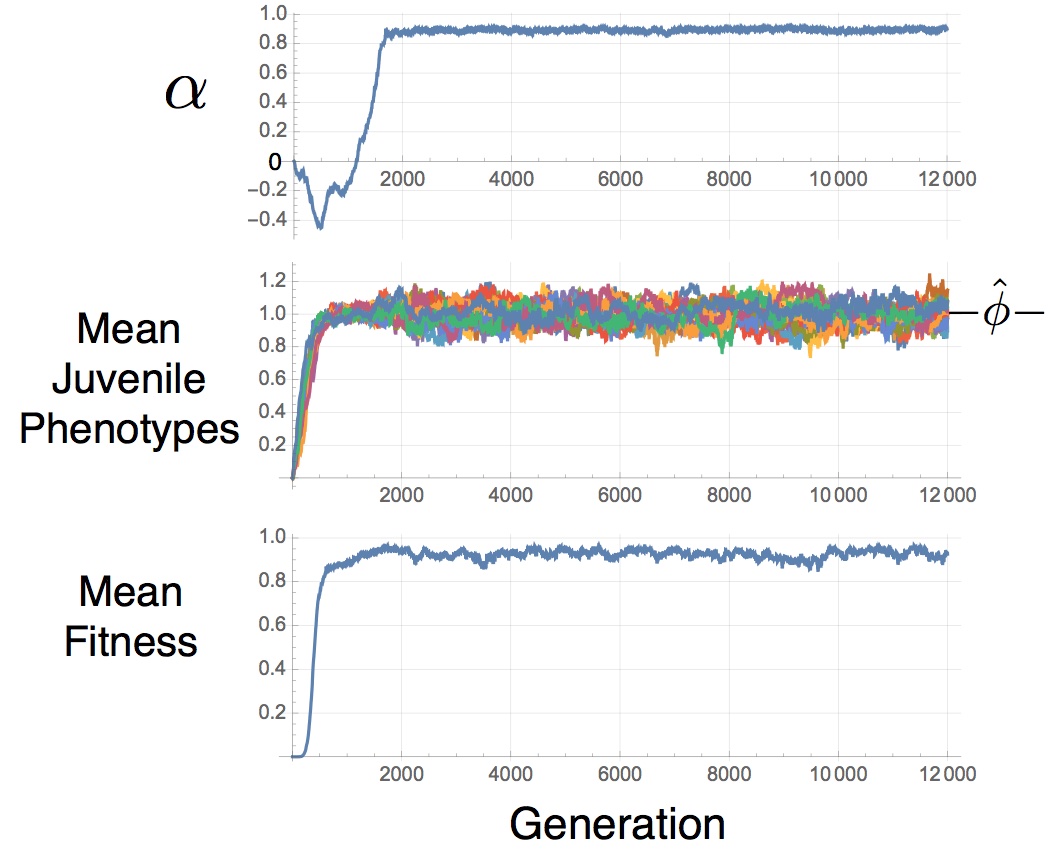}} 
\subcaption{$\alpha \leq 0.95$\label{fig:AlphaLimit24}}
\end{subfigure}
\caption{{\bf An extended evolutionary run with 16 phenotypic variables.} (a) $\alpha$ is allowed to mutate into the hyper-imitation range. (b) $\alpha$ is kept below $0.95$. In both (a) and (b), the top graph shows the evolution of $\alpha$; middle shows the evolution of the population means of each of the 16 variables $\xb_1(t), \ldots, \xb_{16}(t)$; and the bottom shows the mean fitness $\wb(t)$.  Other parameters: $\phih = 1$, $N=1024$, $\sigma^2= 0.01$, $D=16$.}
\end{figure}

This evolution into the hyper-imitation range followed by a collapse of $\alpha$ becomes even more apparent in a model with a multivariate phenotype.  As an illustration, we extend the phenotype to 16 dimensions, $\x = (x_1, \ldots, x_{16})\tr$ and run evolution for 12,000 generations.  A run is shown in Figure \ref {fig:AlphaEvolves16}.  The top graph shows the trajectory of $\alpha$: we see 11 repetitions of the pattern that $\alpha$ evolves into the hyper-imitation range followed by a dramatic evolutionary collapse of $\alpha$ to the middle of the imitation range.  

The middle graph in Figure \ref {fig:AlphaEvolves16} shows the evolution of population means of the 16 juvenile phenotypic variates.  When $\alpha$ enters the hyper-imitation zone, the juvenile phenotypes start diverging from their optimal value (here we set $\phih =1.0$ to shorten the time to reach stabilizing selection).  This is a consequence of the ``evolutionary pathology'' that juvenile phenotypes evolve away from the optimum when $\alpha > 1$.  As a consequence, the mean fitness of the population evolves lower during this phase (bottom graph).  At some point, the random changes in frequency due to Wright-Fisher finite population sampling cause the mean phenotypes to diverge sufficiently from $\phih$ that population effectively returns to the regime of directional selection.  At this point, $\alpha$ is strongly selected to decrease, producing the collapse of $\alpha$ into the mid-imitation range.  Subsequently, the phenotypes evolve back toward $\phih$ and stabilizing selection returns, causing $\alpha$ to evolve larger.  Finally, $\alpha$ again enters the hyper-imitation zone and the cycle repeats.

The collapse phenomenon can be shown to require that $\alpha$ enter the hyper-imitation range by re-running the evolutionary dynamics where an upper limit is placed on the possible values of $\alpha$ below $1.0$.  An evolutionary run is shown in Figure \ref{fig:AlphaLimit24} with the limit $\alpha \leq 0.95$.  After the population enters the stabilizing selection phase, it remains there, with continual limited fluctuations in $\alpha$, $\x$, and $\wb$ due to finite Wright-Fisher sampling.

\subsubsection{Purifying Selection Model}\label{subsub:EvolvesPurifying}

In the purifying selection model, genetic variation for the juvenile phenotype ranges from $0$ to $1$, while the phenotypic optimum is at $1$.  Thus all the variation falls on one side of the optimum.  Runs with fixed $\alpha$ in the previous section revealed in Figure \ref {fig:FixedAlphaPurPhene} that smaller values of $\alpha$ both speed up evolution and produce higher mean fitnesses at mutation-selection balance.  This would lead us to expect that when $\alpha$ is controlled by the modifier locus, $\alpha$ would evolve to decrease.  This is indeed what is observed.
\begin{figure}[H]
\centerline{\includegraphics[width=4in]{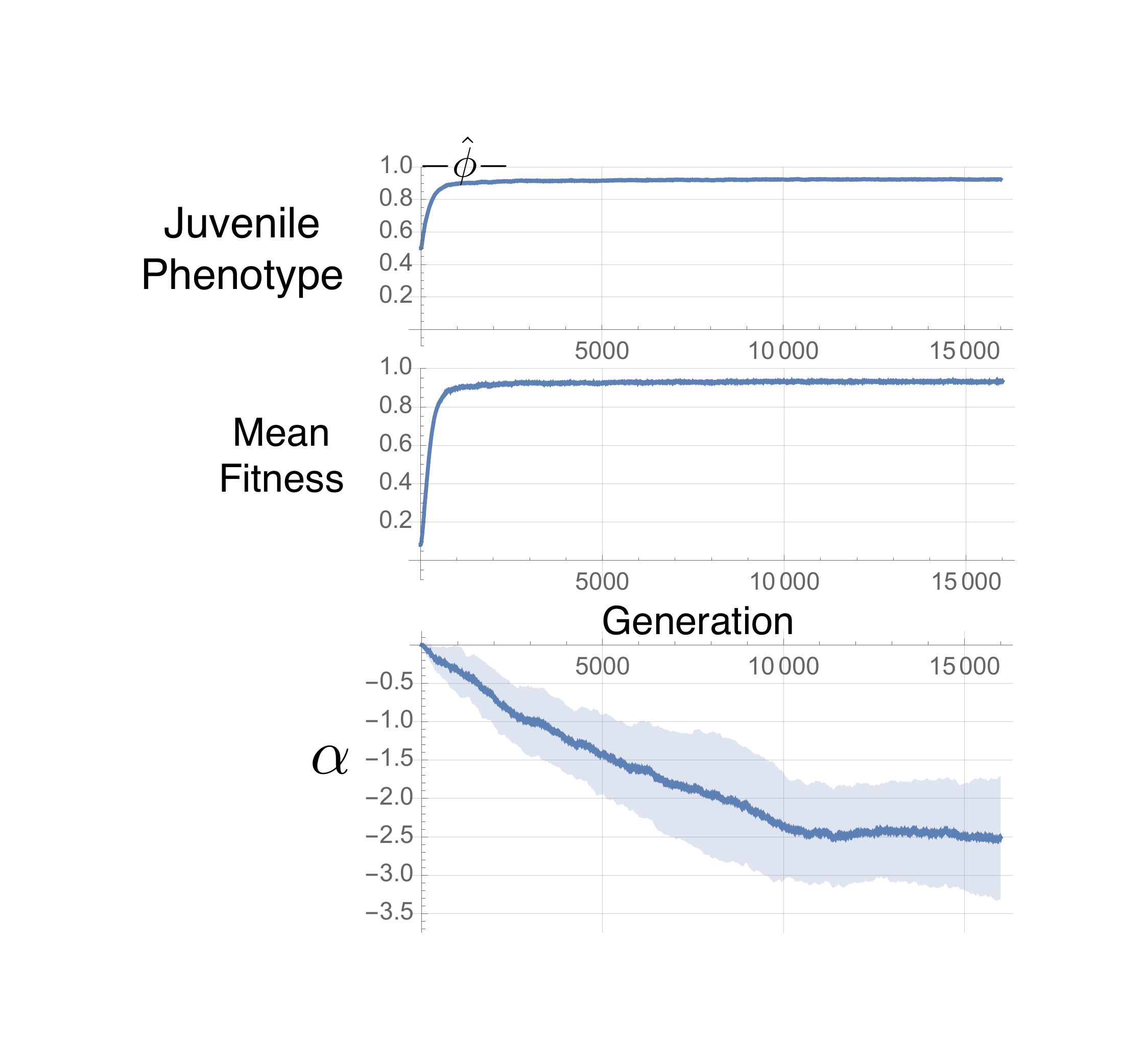}} 
\caption{Average of 24 evolutionary trajectories for the purifying selection model. Shown are the population average juvenile phenotypes (top), mean fitness (middle), and imitation parameter $\alpha$ (bottom).  Long after mutation-selection balance is achieved in the juvenile phenotype, $\alpha$ continues to evolve deep into the anti-imitation range.  Shaded bands are $\pm$ one standard deviation.  Other parameters: $\phih = 1$, $N=1024$, $\beta=10$, $\sigma^2= 0.01$, $L=128$.\label{fig:AlphaEvolvesPur24}}
\end{figure}
Figure \ref {fig:AlphaEvolvesPur24} shows the evolution of the population mean $\alpha$ values averaged over 24 independent evolutionary runs (bottom graph).  Populations are initialized with major genotypes having 64 0-alleles and 64 1-alleles to give $x(0) = 0.5$, and with modifier alleles set to $\alpha(0) = 0$.  The simultaneous values of mean juvenile phenotype $\xb(t)$ and mean fitness $\wb(t)$ and plotted in the top and middle graphs. While each evolutionary trajectory in $\alpha$ is highly meandering, the trajectories show a secular decline in $\alpha$, continuing far into the anti-imitation range long after the phenotype and mean fitness have reached a mutation-selection balance.  

How can this evolution of anti-imitation under purifying selection be understood?  Because the optimal phenotype is extreme, there is no way for the population to surround the optimal phenotype as it does in the stabilizing selection model, where the population mean at mutation-selection balance is closer to the optimum than is the typical individual.  Mutation pressure in the biallelic multilocus model pushes the population very strongly away from the optimal genotype $(1, 1, \ldots, 1)$.  Thus the population will enter a perpetual state similar to directional selection, where anti-imitation moves fitter individuals' adult phenotypes toward the optimum, while moving less fit individuals' adult phenotypes away from the optimum.  The net effect is that the advantage from boosting the fitness of above average individuals outweighs the cost of reducing the fitness of below average individuals---the ``Matthew effect'' \citep{Merton:1968:Matthew} again.

\section{Discussion}

The main general finding of this study is to confirm the value of ``genes as cues'' \citep{Dall:McNamara:and:Leimar:2015:Genes}---in this case, the cues being genes of other organisms.  We show that information on the distribution of genetically determined phenotypes in a population is actionable information:  by organisms becoming ``population geneticists'' and using this information in the right way to plastically change their phenotypes, this information can give it a selective advantage, and genetic variation for organismal capabilities that use this information can come under strong selection.

On the flip side, use of information by the wrong phenotypic plasticity function can slow adaptive evolution, increase genetic load, or even produce an evolutionary pathology where phenotypes evolve in the wrong direction and fitness decreases in evolution.  

When modifier genes are introduced that produce variation in a space of phenotypic plasticity functions, these functions will evolve to utilize this population information to produce adaptive phenotypic plasticity.  The adaptive phenotypic plasticity can evolve to such an extreme, however, that evolution ventures into evolutionary pathology with the evolution of hyper-imitation followed by imitation collapse.

Some points should be emphasized.  None of the phenomena observed here involve alteration of the mutation process.  There is never any ``directed mutation'' in this model.  Mutation remains the same Gaussian normal random variable throughout all the simulations.  Moreover, the ``direction'' that directed variation would require is also entirely missing from this model:  the population genetic information available to the organism never tells it where the phenotypic optimum is;  all the organism knows about is its own juvenile phenotype and the mean juvenile phenotype of the population.  It has no discernment as to what phenotypes should be advantageous.  Even with a lack of this direct information about the phenotypic optimum, however, the knowledge of the population mean phenotype combined with the right phenotypic plasticity function enables the lineage to accelerate its evolution or maximize its mean fitness.  

It should also be noted that in the models examined here, there was always strong selection for either imitation or anti-imitation; the absence of phenotypic plasticity ($\alpha = 0$) was never an evolutionarily stable strategy under any condition of the population.  

We now discuss several aspects of the study.

\subsection{Biological Reality}  
This model of phenotypic plasticity and the behavior it exhibits are novel on a number of fronts.  We are aware of no previous work that investigates anti-imitation, i.e.\ the adoption of traits in the direction opposite to that of a model.  

Every other component of our model, however, when considered separately, has appeared in many models in the literature, these being: natural selection on a quantitative trait, phenotypic optima, Gaussian selection, phenotypic plasticity, imitation, horizontal transmission, Wright-Fisher sampling, modifier genes, and semelparous organisms with discrete non-overlapping generations that never interact with their ancestors.  Each of these common components is in turn based on widely observed empirical biological phenomena.   

But when combined all together, we observe phenomena that are qualitatively novel, and for which we know of no empirical documentation, namely: evolution of imitation, anti-imitation, and hyper-imitation in the absence of cultural transmission, determined by the extent of directional, stabilizing, or purifying selection, and evolutionary pathologies with the repeated buildup and collapse of imitation.  

What are we to make of the novelty of these phenomena?   We propose two principal interpretations:
\begin{enumerate}
\item Since these phenomena have never been documented in nature, the model is invalidated. 
\item Since imitation has been previously thought of only as a phenomenon attending species with social learning, and the usefulness of anti-imitation never imagined, hypothesis-driven research has never considered these phenomena as hypotheses, and so they could have been entirely overlooked.  But with these phenomena as new search-images, perhaps they can become documented. 
\end{enumerate}
We leave it as an open question which of these may be correct.  It may simply be that in nature, the complete set of mechanisms we model never appear together in a single species.  But there are other instances where empirical phenomena had never been investigated until they were postulated theoretically, some phenomena quite illustrious.  The possibility that some structures have been overlooked even in the most thoroughly studied organisms is illustrated by the recent discovery of new organs in humans --- the central nervous system lymphatic vessels \citep{Louveau:etal:2015:Structural},  plantaris ligamentous tendon \citep{Olewnik:etal:2021:Impact}, and tubarial salivary glands \citep{Valstar:etal:2021:Tubarial} --- and in plants:  the discovery of an entirely unknown, environmentally-induced plastic organ, the cantil, in heavily studied \emph{Arabidopsis thaliana} \citep{Gookin:and:Assmann:2021:Cantil}.  Phenomena can be overlooked.

Given the common nature of each of the elements of the models we investigate here, we would expect that the evolutionary dynamics exhibited by their combination would be at least partly manifest in some species.  We hope that researchers of phenotypic plasticity and animal behavior might reexamine their knowledge base in light of these phenomena.

\subsection{Plasticity-First Evolution}  
The hypothesis of ``plasticity-first'' or ``plasticity-led'' evolution proposed by \citet{West-Eberhard:2003:Developmental} continues to attract interest \citep{Levis:and:Pfennig:2016:Evaluating}, and one may ask whether the model and phenomena found here provide examples of it (Justin Yeh, personal communication).  The narrative of plasticity-first evolution is that a novel environment elicits novel phenotypes throughout a population due to phenotypic plasticity, and this is followed by natural selection on any genetic variation that optimizes the form and expression of the novel phenotypes. 

A change in the environment is represented in our models by the population starting out far from the phenotypic optimum.  Because the phenotype in our model is an unstructured quantitative trait, the only option for novelty is its magnitude.  We could therefore consider an adult trait as ``novel'' when it differs from the genetically determined juvenile phenotype, which happens with any nonzero value of imitation parameter $\alpha$.  

But the cause of these ``novel'' adult phenotypes is not environmental change but imitation (or anti-imitation as the case may be).  That is, the adult phenotype $\phi = (1-\alpha) x + \alpha \, \xh$ is a new value different from $x$ due not to a change in the phenotypic optimum $\phih$ (environmental change), but due to phenotypic plasticity being sensitive to the standing genetic variation in the population ($\alpha \neq 0$).  Thus changes in the environment, $\phih$, do not affect phenotypic \emph{expression} but only alter fitnesses.  The production of novel values of $\phi$ under the action of $\alpha$ is solely the result of the genetic variation for the juvenile phenotype present in the population, with no relation to $\phih$.  So we must conclude that our model is not an example of plasticity-first evolution as defined by its authors.

\subsection{Interaction with Cultural Transmission}
For the purpose of investigating imitation without cultural transmission, we construct the life cycle and demography to hermetically seal off one generation from the next.  However,  the phenomena we document could also influence the dynamics of populations that do have cultural transmission.  The phenotypic plasticity in our models, based on organisms observing their conspecifics, can have the same underlying perceptual and behavioral mechanisms as those that would be involved in cultural transmission.  

Once an species has evolved the capability of social learning, or more generally the capability to obtain sensory signals from the phenotypes of conspecifics and to incorporate these signals into plastic behavior or development, the dynamics we document can come into play.  In other words, we know of many species that engage in imitation and social learning.  There is no \emph{a priori} reason these sensory and cognitive capabilities could not tune in to genetically determined juvenile phenotypic variation.  In this case, the selective advantage of anti-imitation during directional selection and imitation during stabilizing selection could operate even when individuals from the previous generation are available to provide information.  More complex models in which organisms can incorporate multigenerational information into their plasticity function merit investigation.  A paradigm for such further study would be the approach taken by \citet{Kuijper:etal:2021:Evolution}, in which they combine into one model the processes of vertical and horizontal learning, prestige and conformity biases, epigenetic inheritance, individual learning, environmental and maternal effects, bet-hedging, genetic cues, and fluctuating environments, for the purpose of determining which processes most dominate the evolutionary dynamics.

\subsection{Heritability}  
Measures of heritability partition phenotypic variance into genetic and environmental terms.  The environmental variance is presumed to be caused by different environments experienced by each organism.  In our models, the environment as represented by the phenotypic optimum, $\phih$, is always kept constant in time, so all organisms experience the same external environment.  If, however, we consider an organism's cohort as part of its environment, then this environment changes constantly throughout evolution.  In this case we can interpret the plastic change from juvenile to adult phenotype, $x$ to $\phi$, as an effect of the environment (the organism's cohort), and use this to calculate the broad sense heritability of the population, which is just the reciprocal of the variance ratio \eqref{eq:VarRatio}:
\hide{
H^2 &= \frac{\Var[x]}{\Var[\phi]}
= \frac {\Var[x]}{\Var[(1-\alpha) x + \alpha \xh]}
}
\ab{
H^2 &= \frac{\Var[x]}{\Var[\phi]}
= \frac{1}{(1-\alpha)^2 }.
}
This expression for the broad sense heritability $H^2$ is less than one (the normal state of affairs) only for $\alpha < 0$ (anti-imitation) or $\alpha> 2$ (hyper-imitation), while for the imitation range, $0 < \alpha < 1$, heritability is greater than $1$, as well as for the low hyper-imitation range $1 < \alpha < 2$.  Heritability becomes infinite at genotype-phenotype disengagement ($\alpha=1$) as there is no variation in the adult phenotype.  Heritability values greater than one occur with imitation because the environmental effect (regression toward the genotypic mean $\xb$) is negatively correlated with the genotypic values.

If a quantitative breeder were trying to measure heritability of a trait, and was unaware that the adult phenotypes were the plastic products of imitation, then the shift in the mean of the offspring compared to the mean of the selected parents due to selective breeding would be incongruously large.  Conversely, if anti-imitation were shaping the adult phenotypes, selective breeding would produce an anomalously small response in the juvenile phenotypes from what was expected based on the adult phenotypes.

\subsection{Numerical Exploration}  
Our emphasis here has been to present the broad qualitative behaviors of the imitation model.  For this reason we have not pursued exhaustive exploration of the parameter space.  We use population size $N=1024$ throughout.  This value is small enough to manifest the stochastic phenomena due to Wright-Fisher sampling, while large enough that natural selection will not be swamped out by stochastic sampling.  We have chosen phenotypic optima $\phih =10$ or $\phih =1$ and mutation intensity $\sigma^2=0.01$ or $\mu = 0.1/L$ that allow both directional and stabilizing or purifying selection phases within the number of generations that we ran.    With a great deal more simulation it would be possible to quantify relationships between the parameters and such population properties as speed of evolution, equilibrium genetic load, phenotypic variance, etc., but such refined relationships are beyond the goals of the present work.

\subsection{Related Models in Evolutionary Computation} 
The idea that organisms can get information on the location of an optimal phenotype by sampling the phenotypes of the population is the foundation for the widely used heuristic optimization method of \emph{Differential Evolution} (DE) \citep{Storn:1996:Usage}.  While in our model mutation is a random variable sampled from a fixed Gaussian distribution, in Differential Evolution the mutation distribution varies as a function of samples drawn from the population.  

A mutation distribution typically used in DE is as follows: an individual $\y_1 \in \Reals^D$ is mutated by adding a random variable $\xi$ where each coordinate $\xi_i$, $i=1, \ldots, D$, is randomly chosen to be either $0$, or the $i$th coordinate of a vector $\z \eqdef \y_j - \y_1 + \alpha(\y_k - \y_h)$, where the other three distinct individuals $j, k, h \neq 1$ are randomly sampled from the population  \citep{Li:and:Li:2019:Differential}.  

In Differential Evolution, therefore, the distribution of genotypes in the population is utilized by the mutation operator itself, not as in our model to alter solely a non-inherited adult phenotype.  This use of population information to alter the mutation mechanism makes Differential Evolution ``Lamarckian'' in the terminology of memetic algorithms \citep{Whitley:etal:1994:Lamarckian}.  Whereas the model we consider would be termed ``Baldwinian'', in that the plastic adult phenotype is not inherited, and the mutation process is not influenced at all by the distribution of organisms in the population.

It remains to be explored whether the imitation/anti-imitation phenotypic plasticity function used here under the control of a modifier gene (``self-adaptation'' in the terminology of \emph{evolution strategies} \citep{Rechenberg:1973}) could be of benefit for evolutionary algorithms, either in the Baldwinian form used here, or in Lamarckian form where adult phenotypes are inherited by the next generation.

\subsection{The Phenotypic Plasticity Function}  
Our choice of phenotypic plasticity function, $\phi = (1-\alpha) x + \alpha \xb$, was crafted to be maximally simple and interpretable as an implementation of imitation.  It assumes that juvenile and adult phenotypes are for the same trait, and even keeps the population mean phenotype unchanged from juvenile to adult, $\overline{\phi} = \xb$. 

But these assumptions are not fundamental to the phenomena identified here.  Since natural selection acts only on the adult phenotype $\phi$, not $x$, there is no reason that $\phi$ and $x$ even need to be within the same metric trait.  The phenotype $x$ could be for a trait entirely unrelated to the adult phenotype $\phi$ except that $x$ is utilized as a cue by the developmental system, much as phenotypic plasticity is usually thought to utilize environmental cues. The phenotypic cue $x$ also need not have been an ancestral part of the developmental pathway for the adult phenotype $\phi$, but could have recently become incorporated into the signaling network for the development of $\phi$.  

In our models, it took hundreds of generations for the phenotypic plasticity function to switch from the anti-imitation evolved during the directional selection to the strong imitation selected for during the stabilizing selection phase.  The mutational variance of $\alpha$ throughout this study was set to $\sigma^2= 0.01$.   A shorter time to adaptive response of $\alpha$ could be achieved by a higher mutational variance for $\alpha$.  Such higher variance could be achieved through epigenetic inheritance.  Further exploration of the mutational process for the imitation parameter is merited.

However, another potential way to shorten the time to adaptive response is if the phenotypic plasticity function were sophisticated enough to detect, solely through population sampling, when the population changed from the directional selection phase to the stabilizing selection phase.  The phenotypic plasticity function here took only a single datum on the population distribution of juvenile phenotypes.  A question for further exploration is whether more detailed sample information can be utilized for to detect the change from directional to stabilizing selection and instantly change plasticity from anti-imitation to imitation.  The necessary information could potentially be contained in the higher moments of the sample distribution of $x$ (skewness, kurtosis, hyperskewness, hypertailedness, etc.), and also in the history of samples taken by successive generations of the organism, which could be stored epigenetically.  A rich space of variation in the plasticity functions could be implemented using genetic programming \citep{Zhang:and:Ciesielski:1999:Genetic} or neural network models \citep{Stanley:2007:Compositional}.

\subsection{Evolutionary Pathologies}  
Hyper-imitation produces the bizarre situation in which organisms have a selective advantage when their juvenile phenotypes are further from the phenotypic optimum than the population mean, which produces the outcome (with fixed $\alpha$) that evolution takes the population away from the phenotypic optimum, and the mean fitness of the population plummets.  When $\alpha$ is free to evolve, evolution for increasing imitation during stabilizing selection gives rise to these pathological hyper-imitators.  This creates an instability during stabilizing selection so that fluctuations of the population mean away from the optimum get amplified, with the resultant drop in mean fitness and collapse of imitation.  These phenomena count as one more addition to the gallery of phenomena caused by frequency-dependent natural selection, which include dynamical chaos \citep{Altenberg:1991} and mean fitness decrease \citep{Asmussen:etal:2004:FDS}.  The repeated epochs of gradual increase in $\alpha$ followed by evolutionary collapse are qualitatively reminiscent of the demographic growth and collapse found by Rogers et al.\ in a model for the evolution of social stratification \citep{Rogers:etal:2011:Spread}.  The dynamics in that model are on their face entirely different from those here, but perhaps there are deeper structural homologies to be elucidated.

The situation in the stabilizing selection model where the evolved mean fitness increases as $\alpha$ approaches $1$ from below, but becomes pathological when $\alpha$ slightly exceeds $1$, could be considered a kind of  ``cliff-edged fitness function'' \citep{Nesse:2004:Cliff}.  The concept was introduced by Nesse as a potential explanation for the evolutionary persistence of diseases like schizophrenia:  selection pushes some phenotypic traits up to the edge of a cliff in phenotype space, like Half-Dome in Yosemite, to a point at the edge where further phenotypic variation becomes very deleterious.  In the models here $\alpha$ does not directly map to fitness, but is rather a parameter in the evolutionary dynamics.  So it could more accurately be described as an example of ``cliff-edged dynamics''.  
 
We found that this instability is prevented by genetic constraints that prevent the production of hyper-imitation.  This is a situation therefore that is a prime candidate for the evolution of \emph{evolvability suppression}.  This idea, first published by \citet{Nunney:1999:Lineage:benefit} and further explored in \citet{Altenberg:2005:Evolvability}, is that when genetic variation has an immediate selective advantage, but produces evolution in a direction that produces long-term harm to the population (an \emph{evolutionary pathology}), then under certain circumstances, modifiers that suppress the generation of this variation will selected for, and can come to prevent the evolutionary pathology.

We would expect that introduction of an additional modifier to the current model that shapes the distribution of mutation effects on the $\alpha$ modifier would evolve so as to keep $\alpha$ less than one.  

\subsection{Purifying versus Stabilizing Selection}  
Very different outcomes were observed between the two models, stabilizing and purifying selection.  In stabilizing selection, the juvenile phenotypes of the population surround the phenotypic optimum when mutation-selection balance has been reached, thereby making the population mean closer to the optimum than the typical individual.  This creates a great advantage to imitating the population mean.  

In purifying selection, the optimal phenotype is at the extreme of what genetic variation is capable of producing, and mutational pressure pushes the population mean away from this optimum.  The juvenile phenotypes are thus incapable of surrounding the phenotypic optimum.  For the fitter part of the population, the population mean is in the opposite direction from the optimum.  This makes anti-imitation an advantage for the fitter organisms, and a disadvantage for the less fit.  But it is this fitter part of the population that contribute most to the future population, so it is advantageous for a plasticity function to adopt anti-imitation and push the adult phenotypes closer to the extreme optimal phenotype.  

Thus imitation never evolves in the purifying selection model, either during directional selection phase or as the population reaches mutation-selection balance.  There are no cliff-edged dynamics, nor emerging evolutionary pathologies, nor instability.

The model of cultural transmission studied by \citet{Gonzalez:Watson:and:Bullock:2017:Minimally} is also of purifying selection.  But in their model, juveniles are produced one by one rather than in a cohort, and they imitate only adult phenotypes, which may themselves have been imitated from earlier generations.  Under those circumstances, imitation is found to evolve under a wide range of population parameters.  Their model does not include the potential for anti-imitation, but its inclusion would be unlikely to change any outcomes because the targets of imitation --- typical adult phenotypes --- are likely to be better than default phenotype of the typical offspring.  This shows how the demographic structure of the population can be determinative in how imitative phenotypic plasticity evolves.

\subsection{Further Exploration}  The models studied here are kept at their most simple so that the relationships between the elements of the model and their behavior are as clear as possible.  But there are many obvious elaborations that can be made which bring in additional biological phenomena.  
\begin{enumerate}
\item Organisms here are all asexual and haploid.  Sexual reproduction, diploid, and genetic recombination all merit additional treatments.  

\item The populations here are unstructured, and only a single population statistic, the mean juvenile phenotype, is extracted by organisms to use to plastically modify their adult phenotypes.  Population subdivision and kin recognition would alter the information used by the organism.   Genetic variation in the choice of individuals from which to collect phenotypic information could come under selection.  One can ask whether related individuals would come to be preferred or avoided as sources of population information to determine the plastic adult phenotype.

\item Selection here does not vary in time.  If the phenotypic optimum changes in time, then the population will be moved away from stabilizing selection to different stages of directional selection.  The most notable phenomenon demonstrated here is that stabilizing selection favors imitation while directional selection favors anti-imitation.  We would expect therefore that different temporal patterns of change in phenotypic optima would produce shifts toward anti-imitation to varying degrees.

\item Mutation rates explored here put the populations in the ``strong mutation'' regime, which keeps the populations polymorphic.  In a ``weak mutation'' regime where the population is monomorphic most of the time, there would be no genetic variation from which the plasticity function could draw, so the adult phenotype would be identical to the juvenile phenotype during each monomorphic epoch, and there would be no selection on $\alpha$.  However, during a selective sweep the population would be polymorphic,  and the phenotypic plasticity function would manifest itself.  Since the invading mutant would benefit from the magnification under anti-imitation of its phenotypic difference from the inferior wild-type, we expect that anti-imitation would again be favored during positive selection in the weak-mutation regime, but this needs to be specifically modeled.

\end{enumerate}

\section{Conclusions}

Here we have shown in a simple model that information on the distribution of genetically determined phenotypes in a population is usable by a species.  When properly incorporated into an organism's phenotypic plasticity, this population genetic information can speed adaptive evolution or reduce genetic load.  There is thus selective opportunity for the organismal capabilities of imitation and anti-imitation.  Cultural transmission of information between generations is not necessary for the evolution of imitation or anti-imitation.  

No special assumptions are required to realize this selective potential beyond an organismal ability to perceive the average genetically determined phenotype for the population.  The novelty of these phenomena raises the question of whether the conditions in the model never occur in nature, or whether these phenomena have simply never been investigated as empirical hypotheses.  This report is only an initial exploration of this space of models.  The results reported here are a novel demonstration of the existence of selection potential for the ``organism as population geneticist''.

\bibliographystyle{humanbio}	

\end{document}